\documentclass[times,final]{elsarticle}

\usepackage{jcomp}
\usepackage{framed,multirow}
\usepackage{todonotes}
\usepackage{comment}

\usepackage{amssymb}
\usepackage{latexsym}
\usepackage{sansmath}

\usepackage{url}
\usepackage{xcolor}
\definecolor{newcolor}{rgb}{.8,.349,.1}

\usepackage{comment}

\usepackage{import}

\usepackage[utf8]{luainputenc}

\usepackage[T1]{fontenc}

\usepackage{graphicx}

\usepackage[english]{babel}
\addto\captionsenglish{}
\addto\captionsenglish{}
\usepackage{csquotes}

\usepackage{newtxtext}
\usepackage{amsthm}
\usepackage[slantedGreek]{newtxmath}
\usepackage[OMLmathsfit]{isomath}
\DeclareMathAlphabet{\mathbfsf}{\encodingdefault}{\sfdefault}{bx}{n}
\usepackage{bm}
\usepackage{envmath}
\usepackage{mathtools}
\usepackage{commath}
\usepackage{siunitx}
\usepackage{IEEEtrantools}

\usepackage[caption=false,font=footnotesize]{subfig}

\usepackage{booktabs}
\usepackage{footmisc}  

\usepackage{url}

\theoremstyle{definition}

\theoremstyle{plain}

\theoremstyle{remark}

\usepackage{lineno}
\modulolinenumbers[5]
\usepackage{umoline}

\usepackage{pgfplots}
\usepackage{pgfplotstable}
\pgfplotsset{compat=newest}
\pgfplotsset{plot coordinates/math parser=false}
\newlength\figureheight
\newlength\figurewidth
\pgfplotsset{every axis plot/.append style={line width=1.5pt},
    legend style={font=\footnotesize, 
        text height=1.0ex,
        draw=black,
        fill=white,
        legend cell align=left}}

\usepackage{hyperref} 
\usepackage[english]{cleveref}

\Crefname{defn}{definition}{definitions}
\Crefname{defn}{Definition}{Definitions}

\Crefname{asm}{assumption}{assumptions}
\Crefname{asm}{Assumption}{Assumptions}

\crefname{lem}{lemma}{lemmas} 
\Crefname{lem}{Lemma}{Lemmas}

\crefname{prop}{proposition}{propositions} 
\Crefname{prop}{Proposition}{Propositions}

\crefname{thm}{theorem}{theorms} 
\Crefname{thm}{Theorem}{Theorms}

\crefname{cor}{corollary}{corollaries}
\Crefname{cor}{Corollary}{Corollaries}
\newcounter{subequation}
\newlength\mtabskip\mtabskip=-1.25cm

\def\mtabLong{long}

\newcommand{\mr}{\mathrm}

\newcommand{\veg}[1]{\bm{#1}}     
\newcommand{\mat}[1]{\mathsfbfit{#1}} 
\renewcommand{\vec}[1]{\mathsfbfit{#1}} 
\newcommand{\op}[1]{\mathcal{#1}} 
\newcommand{\vecop}[1]{\bm{\mathcal{#1}}} 

\renewcommand{\Re}{\mathrm{Re}\;}
\renewcommand{\Im}{\mathrm{Im}\;}

\newcommand{\e}{\mathrm{e}}

\newcommand{\T}{\mr{T}}

\newcommand\restr[2]{{
        \left.\kern-\nulldelimiterspace 
        #1 
        \vphantom{|} 
        \right|_{#2} 
}}

\newcommand\rst[3]{{
        \left.\kern-\nulldelimiterspace 
        #1 
        \vphantom{|} 
        \right|_{#2}^{#3} 
}}

\usepackage{acro}

\DeclareAcronym{DG}
{
    short = DG ,
    long = discontinuous Galerkin
}

\DeclareAcronym{ACA}
{
    short = ACA ,
    long = adaptive cross approximation
}

\DeclareAcronym{EFIE}
{
    short =  EFIE ,
    long = electric field integral equation
}

\DeclareAcronym{MFIE}
{
    short =  MFIE ,
    long = magnetic field integral equation
}

\DeclareAcronym{CFIE}
{
    short =  CFIE ,
    long = combined field integral equation
}

\DeclareAcronym{MUIE}
{
    short =  MUIE ,
    long = Müller integral equation
}

\DeclareAcronym{PMCHWT}
{
    short =  PMCHWT ,
    long = Poggio-Miller-Chang-Harrington-Wu-Tsai integral equation
}

\DeclareAcronym{SPD}
{
    short =  SPD ,
    long = {symmetric, positive definite}
}

\DeclareAcronym{SPSD}
{
    short =  SPD ,
    long = {symmetric, positive semi-definite}
}

\DeclareAcronym{PEC}
{
    short =  PEC ,
    long = perfectly electrically conducting
}

\DeclareAcronym{RWG}
{
    short = RWG ,
    long = Rao-Wilton-Glisson
} 

\DeclareAcronym{BC}
{
    short = BC ,
    long = Buffa-Christiansen
}

\DeclareAcronym{SVD}
{
    short = SVD ,
    long = singular value decomposition
}

\DeclareAcronym{CG}
{
    short = CG ,
    long = conjugate gradient
} 

\DeclareAcronym{PCG}
{
    short = PCG ,
    long = preconditioned conjugate gradient
} 

\DeclareAcronym{CGS}
{
    short = CGS ,
    long = conjugate gradient squared
}

\DeclareAcronym{CMP}
{
    short = CMP ,
    long = Calderón multiplicative preconditioner
} 

\DeclareAcronym{RFCMP}
{
    short = RF-CMP ,
    long = refinement-free Calderón multiplicative preconditioner
} 

\DeclareAcronym{HPD}
{
    short = HPD ,
    long = {Hermitian, positive definite}
} 

\DeclareAcronym{RHS}
{
    short = RHS ,
    long = {right-hand side}
}

\DeclareAcronym{PW}
{
    short = PW ,
    long = {plane wave}
} 

\DeclareAcronym{HD}
{
    short = HD ,
    long = {Hertzian dipole}
} 

\DeclareAcronym{FF}
{
    short = FF ,
    long = {far-field}
} 

\DeclareAcronym{NF}
{
    short = NF ,
    long = {near-field}
}  

\newcolumntype {n}{c}
\newcolumntype {N}{>{\small}c}
\newcolumntype {L}{>{\small}l}
\newcolumntype {F}{>{\footnotesize}c}
\newcolumntype {v}[1]{>{\raggedright \hspace {0pt}} p {#1}}
\newcolumntype {V}[1]{>{\small \raggedright \hspace {0pt}} p {#1}}
\newcolumntype{d}[1]{>{\DC@{.}{.}{#1}}c<{\DC@end}}

\newcolumntype{R}[1]{%
    >{\begin{turn}{90}\begin{minipage}{#1}\small\raggedright\hspace{0pt}}l%
            <{\end{minipage}\end{turn}}%
}

\NewDocumentCommand{\TA}{o}{
    \IfNoValueTF {#1} {%
        \vecop T_{\kern-2pt\mr{A}}
    }
    {
        \vecop T_{\kern-2pt\mr{A},#1}
    }
}

\NewDocumentCommand{\TPhi}{o}{
    \IfNoValueTF {#1} {%
        \vecop T_{\kern-2pt\Phiup}
    }
    {
        \vecop T_{\kern-2pt\Phiup,#1}
    }
}

\NewDocumentCommand{\matTA}{o}{
    \IfNoValueTF {#1} {%
        \mat T_\mr{A}   
        }
    {
        \mat T_{\mr{A},#1}
    }
}

\NewDocumentCommand{\matTPhi}{o}{
    \IfNoValueTF {#1} {%
        \mat T_\Phiup   
        }
    {
        \mat T_{\Phiup,#1}
    }
}

\NewDocumentCommand{\MSL}{o}{
    \IfNoValueTF {#1} {%
        \veg \Psi_\mr{SL}
        }
    {
        \veg \Psi_{\mr{SL},#1}
    }
}

\NewDocumentCommand{\MDL}{o}{
    \IfNoValueTF {#1} {%
        \veg \Psi_\mr{DL}
        }
    {
        \veg \Psi_{\mr{DL},#1}
    }
}

\NewDocumentCommand{\PA}{o}{
    \IfNoValueTF {#1} {%
        \veg \Psi_\mr{A}
        }
    {
        \veg \Psi_{\mr{A},#1}
    }
}

\NewDocumentCommand{\PPhi}{o}{
    \IfNoValueTF {#1} {%
        \veg \Psi_{\Phiup}
        }
    {
        \veg \Psi_{\Phiup,#1}
    }
}

\journal{Journal of Computational Physics}

\begin{document}

\verso{V. Giunzioni \textit{et al.}}

\begin{frontmatter}

\title{Limitations of Nyquist Criteria in the Discretization of 2D Electromagnetic Integral Equations at High Frequency: Spectral Insights into Pollution Effects}

\author[1]{Viviana \snm{Giunzioni}}

\author[2]{Adrien \snm{Merlini}}

\author[1]{Francesco P. \snm{Andriulli}\corref{cor1}}
\cortext[cor1]{Corresponding author: 
  \ead{francesco.andriulli@polito.it}.}
  
\address[1]{Department of Electronics and Telecommunications, Politecnico di Torino, 10129 Turin, Italy}
\address[2]{Microwaves Department, IMT Atlantique, 29238 Brest, France}

\received{}
\finalform{}
\accepted{}
\availableonline{}
\communicated{V. Giunzioni}

\begin{abstract} 

The use of boundary integral equations in modeling boundary value problems---such as elastic, acoustic, or electromagnetic ones---is well established in the literature and widespread in practical applications. These equations are typically solved numerically using boundary element methods (BEMs), which generally provide accurate and reliable solutions. 

When the frequency of the wave phenomenon under study increases, the discretization of the problem is typically chosen to maintain a fixed number of unknowns per wavelength. Under these conditions, the BEM over finite-dimensional subspaces of piecewise polynomial basis functions is commonly believed to provide a bounded solution accuracy, at least for non-trapping geometries. If proven, this would constitute a significant advantage of the BEM with respect to finite element and finite difference time domain methods, which, in contrast, are affected by numerical pollution that causes the number of unknowns per wavelength required to achieve a prescribed solution accuracy to increase with the frequency.

In this work, we conduct a rigorous spectral analysis of some of the most commonly used boundary integral operators and examine the impact of the BEM discretization on the solution accuracy of widely used integral equations modeling two-dimensional electromagnetic scattering from a perfectly electrically conducting cylinder.
We consider both ill-conditioned and well-conditioned equations, the latter being characterized by solution operators bounded independently of frequency.
Our analysis, which is capable of tracking the effects of BEM discretization on compositions and sums of different operators, reveals a form of pollution that affects, in different measures, equations of both kinds. After elucidating the mechanism by which the BEM discretization impacts accuracy, we propose a solution strategy that can cure the pollution problem thus evidenced.
The defining strength of the proposed theoretical model lies in its capacity to deliver deep insight into the root causes of the phenomenon.

\end{abstract}


\end{frontmatter}

\section{Introduction} 

Numerical methods are widely employed to model non-canonical, realistic, scattering problems involving the Helmholtz equation \cite{harrington1993field,jin2015theory}. 
Finite element strategies consist in approximating the solution as a linear combination of a finite number of basis functions defined on finite elements.

Given $v\in V$ and $V_N$ a finite dimensional subspace of $V$ of dimension $N$, the best approximation of $v$ in $V_N$ is given by the orthogonal projection \cite{steinbach2008numerical,jin2015theory} $P_N:V\rightarrow V_N$ of $v$ onto $V_N$, i.e., 
\begin{equation}
    \|(I-P_N)v\|_{V} = \min_{w_N\in V_N}\|v-w_N\|_V\,.
\end{equation}
When fixing an error threshold $\epsilon$, there exists a minimum number of elements $N(\epsilon)$ such that the best approximating function of $v$ in the corresponding finite element space $V_{N(\epsilon)}$ has an error less than or equal to $\epsilon$,
\begin{equation}
    \|(I-P_{N(\epsilon)})v\|_{V} \le \epsilon\,.
\end{equation}
To achieve the same accuracy, a numerical method requires a number of unknowns $N_{\text{num}}(\epsilon)\ge N(\epsilon)$. If the ratio $N_{\text{num}}(\epsilon)/ N(\epsilon)$ remains bounded as the wavenumber increases, the numerical method is quasi-optimal. In contrast, if the ratio $N_{\text{num}}(\epsilon)/ N(\epsilon)$ increases as the wavenumber increases in the high-frequency limit, we say that the numerical method is affected by pollution \cite{babuska1995generalized}.

From the Weyl law describing the asymptotic behavior of the eigenvalues of the Helmholtz operator \cite{weyl1912asymptotische,gustafsson2025degrees} or, in the one-dimensional case, from the Shannon-Nyquist sampling theorem \cite{shannon1949communication}, we know that the number of piecewise polynomial basis functions required to accurately approximate a function in $\mathbb{C}^d$ oscillating with frequency $\lesssim k$ is approximately $k^d$. This means that the natural growth of the number of degrees of freedom is $N\sim k^d$. Rigorous considerations and error bounds can be found in \cite{graham2015when,galkowski2025lower}.
Hence, a growth in frequency of the parameter $N_{\text{num}}$ higher than $k^d$ denotes pollution.

Some of the most widespread numerical methods, such as the finite element and the finite difference methods (FEM/FDM) are affected by pollution. After the initial observations and analyses of the phenomenon in the eighties \cite{bayliss1985accuracy,aziz1988two}, in-depth investigations during the following decade \cite{ihlenburg1995finite,ihlenburg1995dispersion,babuska1997posteriori} led to a comprehensive understanding of the problem.
In particular, the pollution affecting the 
FEM is related to a phase lag of the numerical solution with respect to the analytic one. In other words, the wavenumber of the numerical solution diverges from the true wavenumber when using a number of finite elements proportional only to $k^d$, resulting in numerical dispersion. Significant investigations have been directed toward the definition of strategies to minimize the computational burden related to the increase of the ratio $N_{\text{num}}(\epsilon)/ N(\epsilon)$ in frequency. While the pollution effect can be avoided in one-dimensional problems \cite{babuska1995generalized,babuska1997pollution,babuska1997partition,wang2014pollutionfree}, it is not the case in higher-dimensional problems \cite{babuska1995generalized,babuska1997pollution}. Numerous techniques have been proposed to control and minimize the pollution error in two and three-dimensional problems. Even though they are capable of significantly mitigating the requirement for bounded accuracy from $hk^2=\text{const.}$ up to $h^3k^{3.5}=\text{const.}$, where $h$ denotes the size of the elements of the mesh, \cite{babuska1995generalized,babuska1997pollution,franca1997residualfree,thompson1995galerkin,ihlenburg1997finite,deraemaeker1999dispersion}, the related numerical dispersion arising at high-frequencies represents a significant drawback of finite element-based simulations.

The boundary element method (BEM) is another popular strategy to numerically solve Helmholtz scattering problems. With respect to finite element methods, it only requires the discretization of the boundary of the scatterer and allows for boundary conditions to be enforced automatically, at the cost of building non-sparse linear systems of equations. In the engineering community, it is common practice to numerically solve BEM Helmholtz scattering problems by fixing a number of degrees of freedom per wavelength, that is, by assembling and solving linear systems of size proportional to $k^{d-1}$, where $d$ is the dimension of the problem \cite{betcke2017computationally,marburg2002six,chaillat2017fast}. This rule of thumb reflects the common belief that the $h-$BEM is less affected by numerical pollution than the $h-$FEM, at least for non-trapping geometries \cite{barucq2017symmetric,galkowski2022does}. Various numerical experiments proposed in the literature suggest the quasi-optimality of some boundary integral equations (BIEs) under certain conditions \cite{betcke2017computationally,barucq2017symmetric}. On the other hand, a form of pollution similar to that observed in the $h-$FEM has been numerically identified by Marburg in some of his experiments \cite{marburg2018pollution,marburg2016numerical}.

Some theoretical investigations resulting in $k$-explicit quasi-optimality statements about boundary integral equations have been proposed in the numerical analysis literature, mainly focused on the direct and indirect second kind formulations for the exterior Dirichlet problems. Fourier analyses performed  on integral operators evaluated over spherical domains \cite{buffa2006acoustic,banjai2007refined} have yielded quasi-optimal, frequency-independent, error estimates for the solution of the BEM discretized standard combined equations for the above problem. Similar results for the same equations applied to scatterers bounded by analytic curves or surfaces have been demonstrated in \cite{lohndorf2011wavenumberexplicit}.
The analysis proposed is based on a novel decomposition of the combined integral operators \cite{melenk2012mapping} which allows for a separate examination of their low- and high-frequency spectral components through Fourier analysis.
Based on a similar idea of splitting of diverse frequency components, the work in \cite{galkowski2022does} finally provides a proof of the fact that the Helmholtz exterior Dirichlet problem does not suffer from the pollution effect when the obstacle is non trapping. The analysis here takes advantage of the concept of semiclassical pseudodifferential operators \cite{zworski2012semiclassical}, which, with respect to the standard, or homogeneous, pseudodifferential operators \cite{hsiao2008boundary}, are particularly tailored to the high-frequency analysis required. Further investigations about these topics by some of the same authors are summarized in the manuscript \cite{galkowskihelmholtz}, expanding the analysis to the Helmholtz exterior Neumann problem solved by means of a Calderón regularization of the standard combined operator, involving the hypersingular operator. One of the main results obtained is that ``If the inverse of the boundary integral operator is not bounded independently of $k$, then Galerkin and collocation method based on piecewise polynomials suffer from the pollution effect'' \cite{galkowskihelmholtz}. Additionally, the analysis is expanded to BEM discretization approaches involving finite dimensional subspaces of trigonometric polynomials, which, on the other hand, provide quasi-optimal error. However, it is worth noting that, when considering products of operators, a simplified model of implementation of the Galerkin approximation method is analyzed, which ignores the issue of discretizing the product of operators as product of matrices discretizing single operators.

In this work, we propose a fundamentally different approach for analysing the problem that is based on the spectral Fourier analysis of operator matrices rather than the operators themselves, and is thus capable of explaining the effects of BEM discretization on integral operators and the resulting solution errors. We focus on the two-dimensional electromagnetic scattering from a perfectly electric conducting (PEC) cylinder, modeled by electric and magnetic field integral equations (EFIEs/MFIEs) discretized with piecewise polynomials.
Within the developed theoretical framework, we derive closed-form expressions for the current and scattering errors resulting from BEM discretization of the formulations under study. Through their high-frequency asymptotic analysis, we identify a form of pollution affecting the TE-EFIE equation, which, to the best of the authors' knowledge, has not been previously reported in the literature.
Additionally, we extend the analysis to a Calderón-stabilized combined field equation (CCFIE), whose solution operator remains bounded independently of frequency. Our modeling techniques, which track and predict the effects of discretizing a product of operators as a product of matrices that discretize individual operators, reveal negative repercussions of the BEM discretization even on the well-behaved CCFIE. 
Finally, and most importantly, the proposed analytical approach offers simple yet rigorous explanations of the pollution mechanism, identifying its root causes. This profound understanding of the issue directly opens the way to solution strategies based on operator filtering techniques \cite{masciocchi2023operator}, as demonstrated in this work through both analytical and numerical results.

This paper is organized as follows: after setting the background and notation in Section~\ref{sec:backgroun}, we introduce in Section~\ref{sec:spectralerror} the spectral error analysis that is the foundation of this work, treating the different operators in the various regions of the spectra and compositions of operators. After moving to the current and scattering error analysis in Sections~\ref{sec:currenterror} and \ref{sec:scatteringerror} that show the existence of a pollution effect in some of the equations considered, we delineate in Section~\ref{sec:filtering} a filtering strategy capable of suppressing the components of the spectral error increasing in frequency and neutralizing the pollution. Numerical results presented in Section~\ref{sec:numresults}
validate the theoretical framework. Preliminary and partial results have been previously presented in the conference contribution \cite{giunzioni2024highfrequency}.

\section{Background and notation}
\label{sec:backgroun}
Consider the time harmonic electromagnetic scattering from a PEC cylinder indefinitely extended along the longitudinal direction $\hat{\veg z}$. Let $\Omega$ be the open set modeling the circular transversal cross-section of the cylinder of radius $a$ and $\Gamma\coloneq\partial \Omega$ be its two-dimensional circular contour. The exterior space $\mathbb{R}^2\backslash\Omega$ is characterized by its impedance $\eta = \sqrt{\mu/\epsilon}$ and the corresponding wavenumber $k = \omega \sqrt{\mu\epsilon}$. By defining the azimuthal angle $\phi \in [0,2\pi)$, $\Gamma$ can be parametrized in $\phi$ in the 2D Cartesian plane as 
$\Gamma = {\left[a\cos(\phi),a\sin(\phi)\right]^\T / \phi \in [0,2\pi)}$.

The single-layer, double-layer, adjoint double-layer, and hypersingular operators are defined as
\begin{align}
    \op S^k f(\veg \rho) &\coloneqq k \int_\Gamma G^k (\veg \rho,\veg \rho^\prime) f(\veg \rho^\prime) d \veg \rho^\prime\,,\\
    \op D^k f(\veg \rho) &\coloneqq \int_\Gamma \frac{\partial}{\partial n^\prime} G^k (\veg \rho,\veg \rho^\prime) f(\veg \rho^\prime) d \veg \rho^\prime\,,\\
    \op D^{*k} f(\veg \rho) &\coloneqq \int_\Gamma \frac{\partial}{\partial n} G^k (\veg \rho,\veg \rho^\prime) f(\veg \rho^\prime) d \veg \rho^\prime\,,\\
    \op N^k f(\veg \rho) &\coloneqq - \frac{1}{k}\frac{\partial}{\partial n} \int_\Gamma \frac{\partial}{\partial n^\prime} G^k (\veg \rho,\veg \rho^\prime) f(\veg \rho^\prime) d \veg \rho^\prime\,.
\end{align}
The two-dimensional free-space Green's function is $G^k(\veg \rho, \veg \rho^\prime) \coloneqq -\frac{\mathrm{j}}{4} H_0^{(2)} (k|\veg \rho- \veg \rho^\prime|)$, where $H_0^{(2)}$ is the Hankel function of the second kind \cite{abramowitz1964handbook}.
These are the building blocks for the electric and magnetic field integral equations, that relate the longitudinal and transversal electric current densities $J_z$ and $J_t$ to the impinging electromagnetic fields $(E_z,H_t)$ and $(E_t,H_z)$ \cite{peterson1998computational,jackson1999classical,chew1995waves}. In the transverse magnetic (TM) polarization, they read
\begin{align}
    \op S^k (J_z) (\veg \rho) &= \frac{E_z(\veg \rho)}{\mathrm{j} \eta} \,,\label{eqn:TMEFIE}\\
    \left(\frac{1}{2} \op I +  \op D^{*k} \right) (J_z) (\veg \rho) &= H_t(\veg \rho) \,,\label{eqn:TMMFIE}
\end{align}
while in the transverse electric (TE) polarization they read
\begin{align}
     \op N^k (J_t) (\veg \rho) &= -\frac{ E_t(\veg \rho)}{\mathrm{j}\eta}\,, \label{eqn:TEEFIE}\\
    \left(\frac{1}{2} \op I -  \op D^k \right) (J_t) (\veg \rho) &= - H_z(\veg \rho)\,.\label{eqn:TEMFIE}
\end{align}

In this work, we will mainly consider the regime in which the number of degrees of freedom used in the discretization of the problem increases proportionally with the frequency, which is equivalent to fixing a constant number of unknowns per wavelength; this setting is often referred to as the high-frequency regime \cite{adrian2021electromagnetic}.

In this regime, the EFIE and MFIE suffer from several sources of instability such as spurious resonances, manifesting themselves as a finite number of eigenvalues of the operators going to zero \cite{mautz1978hfielda,adrian2021electromagnetic}, or the high-frequency breakdown, which is a growth of the condition number as the frequency increases. In particular, the condition number of the matrices discretizing the TM- and TE-EFIE grows in frequency as $\mathcal{O}((ka)^{1/3})$ as $ka\rightarrow \infty$ away from resonances \cite{warnick2000accuracy,darbas2004preconditionneurs}. Both issues can be addressed, for a class of scatterers, by combining and preconditioning the EFIE and MFIE to form the Calder\'{o}n combined field integral equation \cite{andriulli2015high,consoli2022fast}, denoted in the following as ``CCFIE'', that reads for TM and TE polarizations respectively
\begin{align}
    &\left[{\op N^{\tilde{k}}} \op S^k  + \left(\frac{1}{2} \op I -  {\op D}^{\tilde{k}} \right)\left(\frac{1}{2} \op I +  \op D^{*k} \right)\right] (J_z) (\veg \rho) = \frac{\op N^{\tilde{k}}}{\mathrm{j}\eta}E_z(\veg \rho)+\left(\frac{1}{2} \op I -  {\op D}^{\tilde{k}} \right)H_t(\veg \rho)\,,\label{eqn:TMCCFIE}\\
    &\left[{\op S^{\tilde{k}}} \op N^k  + \left(\frac{1}{2} \op I +  {\op D^{*\tilde{k}}} \right)\left(\frac{1}{2} \op I -  \op D^k \right)\right] (J_t) (\veg \rho) =-\frac{\op S^{\tilde{k}}}{\mathrm{j}\eta}E_t(\veg \rho)-\left(\frac{1}{2} \op I +  {\op D^{*\tilde{k}}} \right)H_z(\veg \rho)\,,\label{eqn:TECCFIE}
\end{align}
where $\tilde{k} \coloneqq k-\mathrm{j}0.4k^{1/3}a^{-2/3}$ \cite{antoine2006improved,darbas2006generalized,boubendir2014wellconditioned}. The complex wavenumber employed in the evaluation of the preconditioning operator matrices is shown to be optimal over the circle of radius $a$ and its use allows to define  Calder\'{o}n combined field operators (CCFIOs) which are bounded and whose inverses are bounded in the high-frequency limit \cite{andriulli2015high,consoli2022fast}.

In order to discretize and numerically solve the above integral equations, we consider a uniform discretization of the boundary $\Gamma$ composed of $N$  circular arcs of length $h= 2 \pi a/N$. The vertices of these elements lie at the points $\left[a \cos(\phi_n),a \sin(\phi_n)\right]^\T$, where $\phi_n\coloneqq 2\pi n/N$ for $n=0,\dots,N-1$. In the following, we will assume $N$ odd for simplicity and without loss of generality. We further introduce the parameter $n_\lambda\coloneqq \lfloor2\pi/(k h)\rfloor$, quantifying the ratio between the wavelength and the mesh parameter $h$.

Over this mesh, we define a set of $N$ pyramidal, or Lagrangian, basis functions $\{f_i\}_{i=0}^{N-1}$ attached over the $N$ vertices and having domain $2 h$. In particular, $f_0$ is defined as
\begin{equation}
    f_0(r)=\begin{cases}
        \frac{1}{h}(a\phi_1-r), & \text{if $0\le r < a\phi_1$}\\
        \frac{1}{h}(r-a\phi_{N-1}), & \text{if $a\phi_{N-1}\le r < a(\phi_{N-1}+\frac{2\pi}{N})$}
    \end{cases}
\end{equation}
and $f_i$ is given by $f_i(r) = f_0(r-a\phi_i)$, where we have denoted by $r \coloneqq a\phi$ the curvilinear abscissa. By following the standard Galerkin procedure, we expand the unknowns of the above integral equations as linear combinations of the basis functions $\{f_i\}_{i=0}^{N-1}$ and test the 
resulting equations with test functions of the same kind, resulting in the linear systems
\begin{align}
    \mat S^k \hat{\vec J}_z &= \vec E_z / (\mathrm{j}\eta) \,.\label{eqn:TMEFIEd}\\
    \left(\mat G /2 + \mat D^{*k} \right) \hat{\vec J}_z &= \vec H_t \,.\label{eqn:TMMFIEd}\\
    \mat N^k \hat{\vec J}_t &= -\vec E_t / (\mathrm{j}\eta) \,.\label{eqn:TEEFIEd} \\
    \left(\mat G /2 - \mat D^{k} \right) \hat{\vec J}_t &= -\vec H_z \,.\label{eqn:TEMFIEd}\\
    \mat C^{\text{TM}}\hat{\vec J}_z \coloneq \left[\mat N^{\tilde{k}} \mat G^{-1} \mat S^k + 
    \left(\mat G /2 - \mat D^{\tilde{k}} \right) \mat G^{-1} \left(\mat G /2 + \mat D^{*k} \right)\right]\hat{\vec J}_z &= \mat N^{\tilde{k}} \mat G^{-1}\vec E_z / (\mathrm{j}\eta) +  \left(\mat G /2 - \mat D^{\tilde{k}} \right) \mat G^{-1} \vec H_t \,.\label{eqn:CTM}\\
    \mat C^{\text{TE}}\hat{\vec J}_t \coloneq \left[\mat S^{\tilde{k}} \mat G^{-1} \mat N^k + 
    \left(\mat G /2 + \mat D^{*\tilde{k}} \right) \mat G^{-1} \left(\mat G /2  - \mat D^{k} \right)\right]\hat{\vec J}_t &= -\mat S^{\tilde{k}} \mat G^{-1}\vec E_t / (\mathrm{j}\eta) -  \left(\mat G /2 + \mat D^{*\tilde{k}} \right) \mat G^{-1} \vec H_z \,.\label{eqn:CTE}
\end{align}
where the element $(m,n)$ of a matrix $\mat{O}$, which acts as a placeholder for $\mat S$, $\mat N$, $\mat D$, $\mat D^*$, $\mat G$, is 
\begin{equation}
    \mat{O}_{mn} = \frac{1}{h}\int_0^{2\pi a} \mathrm{d}r \, f_m(r) \,  \left(\op O f_n(r)\right)\,,
\end{equation}
where $\mathcal{O}$ stands for $\mathcal{S}$, $\mathcal{N}$, $\mathcal{D}$, $\mathcal{D}^*$, $\mathcal{I}$ (where $\mathcal{I}$ denotes the identity operator), and the $n$-th element of the array $\vec F$, where $\vec F$ stands for $\vec E_z$, $\vec H_t$, $\vec E_t$, $\vec H_z$, is given by
\begin{equation}
    \vec F_n = \frac{1}{h}\int_0^{2\pi a} \mathrm{d}r \, f_n(r) F(r) \,,\label{eqn:Fn}
\end{equation}
where $F$ stands for $ E_z$, $ H_t$, $ E_t$, $ H_z$. The arrays $\hat{\vec J}_z$ and $\hat{\vec J}_t$ contain the weights of the linear combinations of pyramid basis functions expanding the unknown functions.

Because $\Gamma$ is a circle, one can derive close form expressions for the $q-$th eigenvalue $\lambda_q^{\op O}$ of a continuous operator $\op O$ \cite{hsiao1994error}, which for the operators under study yields 
\begin{align}
    \lambda_q^{\op S^k} &= -\frac{\mathrm{j} k \pi a}{2} J_{q}(k a) H_{q}^{(2)}(k a)\,,
    \label{eqn:lamS}\\
    \lambda_q^{\op D^k} &= \lambda_q^{\op D^{*k}} = -\frac{\mathrm{j} k \pi a}{4} \left[J_{q}(k a) H_{q}^{(2)}(k a) \right]^\prime\,,
    \label{eqn:lamD}\\
    \lambda_q^{\op N^k} &= \frac{\mathrm{j} k \pi a}{2} J^\prime_{q}(k a) H_{q}^{(2)\prime}(k a)\,,
    \label{eqn:lamN}
\end{align}
where $J_{q}(k a)$ and $H_{q}^{(2)}(k a)$ denote the Bessel function of first kind and the Hankel function of second kind of order $q$ and argument $(ka)$.
Note that these functions are linked by the equality $H_{q}^{(2)}(k a)=J_{q}(k a)-\mathrm{j} Y_{q}(k a)$, where $Y_{q}$ is the Bessel function of second kind and order $q$ \cite{abramowitz1964handbook}. 
On the circular domain $\Gamma$ these operators share the same eigenfunctions and thus commute with one another. In particular, the eigenfunction associated with the eigenvalue $\lambda_q^{\op O}$ is the complex exponential $\phi \mapsto \e^{-\mathrm{j} q \phi}$ \cite{hsiao1994error}.
From this commutative property, we can infer that the eigenvalues of the TM/TE-MFIO, $\op I/2+\op D^{*k}$ and $\op I/2-\op D^k$, are \cite{dominguez2007hybrid,abramowitz1964handbook}
\begin{align}
    \lambda_q^{\text{TM-MFIO}^k}&=-\frac{\mathrm{j}\pi k a}{2} J_q^\prime(ka) H_q^{(2)}(ka)\,,\\
    \lambda_q^{\text{TE-MFIO}^k}&=\frac{\mathrm{j}\pi k a}{2} J_q(ka) H_q^{(2)\prime}(ka)\,.
\end{align}

On the particular geometrical and functional setting described above, the eigenvalues of the operator matrices are also known in close form \cite{warnick2001spectrum,davis2005error,warnick2008numerical} and can be expressed as
\begin{equation}
    \hat{\lambda}_q^{\mat O} = \sum_{s=-\infty}^{\infty}\lambda^{\op O}_{(q+sN)}F_{-(q+sN)}F_{(q+sN)}\,,
    \label{eqn:WarnickFormula}
\end{equation}
where $F_q$ represents the $q-$th Fourier coefficient of the function $f_0$ \cite{warnick2000accuracy,warnick2008numerical},
\begin{equation}
    F_q\coloneqq \frac{N}{2\pi}\int \mathrm{d}\phi f_0(\phi)e^{\mathrm{j}q\phi} = \left(\frac{\sin(\pi q/N)}{(\pi q/N)}\right)^{2} \,.
\end{equation}
Equation \eqref{eqn:WarnickFormula} is exact 
if the matrices are obtained in infinite precision and infinite accuracy. Following from the application of the Green's function addition theorem, the analysis leading to the definition of $\hat{\lambda}_q^{\mat O}$ also shows that the eigenvectors of the matrices are shared between different operator matrices and simply correspond to the discretization of the continuous eigenfunctions, $\phi \mapsto e^{-\mathrm{j} q \phi}$, in the chosen functional basis.

The spectral relative difference between continuous and discrete eigenvalues can be derived from the knowledge of both ${\lambda}_q^{\mat O}$ and $\hat{\lambda}_q^{\mat O}$ as
\begin{equation}
    E_q^{\op O} \coloneq \frac{\hat{\lambda}_q^{\mat O}-{\lambda}_q^{\op O}}{{\lambda}_q^{\op O}} = E_q^P+E_q^{A,\op O}\,,
    \label{eqn:spectralerror}
\end{equation}
where
\begin{align}
    E_q^P &\coloneq F_{-q}F_q-1 \,,\label{eqn:projerror}\\ 
    E_q^{A,\op O} &\coloneq \frac{1}{\lambda_q^{\op O}}\sum_{s\ne 0}\lambda^{\op O}_{(q+sN)}F_{-(q+sN)}F_{(q+sN)}
    \label{eqn:aliaserror}
\end{align}
represent the spectral projection and aliasing error contributions, respectively \cite{warnick2000accuracy,warnick2008numerical}.

\section{Spectral error analysis}
\label{sec:spectralerror}

We aim at analyzing the spectral relative error (the relative difference between the eigenvalues of the continuous operators and that of their discrete counterparts) in the high-frequency regime.
The analysis will also include the spectral error of the combinations and products of integral operators, such as the ones appearing in the magnetic field integral operator (MFIO) and in the Calderón combined field integral operator (CCFIO).

\subsection{Sums and products of operators}

Given two matrices $\mat M$ and $\mat N$ that share the same eigenvectors and with eigenvalues $\hat{\lambda}_q^{\op{M}}={\lambda}_q^{\op{M}}(1+E_q^{\op{M}})$ and $\hat{\lambda}_q^{\op{N}}={\lambda}_q^{\op{N}}(1+E_q^{\op{N}})$ (with $E_q^{\op{M}}$ and $E_q^{\op{N}}$ in the form of \eqref{eqn:spectralerror}), the $q-$th eigenvalue of their sum $(\mat M+\mat N)$ is $(\hat{\lambda}_q^{\op{M}}+\hat{\lambda}_q^{\op{N}})$, with relative error with respect to $({\lambda}_q^{\op{M}}+{\lambda}_q^{\op{N}})$ given by
\begin{equation}
E_{q}^{\op{M}+\op{N}}=\frac{(\hat{\lambda}_q^{\op{M}}+\hat{\lambda}_q^{\op{N}})-({\lambda}_q^{\op{M}}+{\lambda}_q^{\op{N}})}{({\lambda}_q^{\op{M}}+{\lambda}_q^{\op{N}})}=E_q^P+\frac{{\lambda}_q^{\op{M}}E_{q}^{A,\op{M}}+{\lambda}_q^{\op{N}}E_{q}^{A,\op{N}}}{{\lambda}_q^{\op{M}}+{\lambda}_q^{\op{N}}}\,,
\label{eqn:opsum}
\end{equation}
in which the second term represents the aliasing error of the sum of the two operators.
The $q-$th eigenvalue of their product $(\mat M \mat N)$ is $\hat{\lambda}_q^{\op{M}}\hat{\lambda}_q^{\op{N}}$, with its relative error with respect to ${\lambda}_q^{\op{M}}{\lambda}_q^{\op{N}}$ given by
\begin{equation}
E_{q}^{\op{M}\cdot\op{N}}=\frac{(\hat{\lambda}_q^{\op{M}}\hat{\lambda}_q^{\op{N}})-({\lambda}_q^{\op{M}}{\lambda}_q^{\op{N}})}{({\lambda}_q^{\op{M}}{\lambda}_q^{\op{N}})}=E_q^{\op{M}}+E_q^{\op{N}}+E_q^{\op{M}} E_q^{\op{N}}\,.
\label{eqn:opprod}
\end{equation}

\subsection{The spectral aliasing error}

Differently from the projection error, the aliasing error is strictly associated to the continuous operator from which the matrix derives. Hence, to understand the high-frequency behavior of the aliasing spectral error of the formulations under test, we need to study the behavior of the eigenvalues of the integral operators in the high-frequency limit. The analysis will be performed in three spectral regions: the hyperbolic region of propagative modes at indices $q \ll ka$, the transition region of modes at indices $q \simeq ka$, and the elliptic region of evanescent modes at indices $q \gg ka$.

\subsubsection{Eigenvalues in the hyperbolic region}
\label{sec:HAnal}
To study the behavior of the eigenvalues of the single-layer, double-layer, and hypersingular operators in the hyperbolic region \cite{buffa2006acoustic} we use the large-argument principal asymptotic expansions of the Bessel and Hankel functions 
\cite[Eq. 9.2.1, 9.2.4]{abramowitz1964handbook} valid for fixed positive indices and growing argument.
It follows that the operator eigenvalues in the hyperbolic region in the high-frequency limit can be approximated as
\begin{align}
    &\lambda_q^{\op S^k} \sim -\mathrm{j} e^{-\mathrm{j} \beta} \cos(\beta)\,,\\
     &\lambda_q^{\op N^k} \sim - \sqrt{\mathrm{j}} e^{-\mathrm{j} \gamma} \sin(\beta)\,,\\
     &\lambda_q^{\text{TM-MFIO}^k} \sim \mathrm{j} \sqrt{\mathrm{j}} e^{-\mathrm{j} \gamma} \sin(\beta)\,,\\
    &\lambda_q^{\text{TE-MFIO}^k} \sim  e^{-\mathrm{j} \beta} \cos(\beta)\,,\\
     &\lambda_q^{\op D^k} \sim -\mathrm{j} e^{-2\mathrm{j}\gamma}/2\,,
\end{align}
where $\beta = ka- q \pi/2-\pi/4$, $\gamma = \beta + \pi/4$, and the symbol $\sim$ indicates an asymptotic equality \cite{abramowitz1964handbook}.
These expressions show that $\lambda_q^{\op S^k}$ and $\lambda_q^{\op N^k}$ have a constant behavior in frequency away form resonances, both in real and imaginary parts. The same holds true for the eigenvalues of the MFIO and of the double-layer operator.

\subsubsection{Eigenvalues in the transition region}
\label{sec:TAnal}
To study the behavior of the eigenvalues of the single-layer, double-layer, and hypersingular operators in the transition region \cite{buffa2006acoustic,warnick2000accuracy} we leverage the large order asymptotic expansions of the Bessel and Hankel functions and of their derivatives in the transition region. The asymptotic expansions \cite[Eq. 9.3.31-9.3.34]{abramowitz1964handbook} read
\begin{align}
    J_q(q) &\sim q^{-1/3} \mu \Sigma_\alpha-q^{-5/3} \nu \Sigma_\beta\,,\\
    Y_q(q) &\sim -q^{-1/3} \sqrt{3}\mu \Sigma_\alpha-q^{-5/3} \sqrt{3}\nu \Sigma_\beta\,,\\
    J_q^\prime(q) &\sim q^{-2/3} \nu \Sigma_\gamma-q^{-4/3} \mu \Sigma_\delta\,,\\
    Y_q^\prime(q) &\sim q^{-2/3} \sqrt{3}\nu \Sigma_\gamma+q^{-4/3} \sqrt{3}\mu \Sigma_\delta\,,
\end{align}
where $\mu$, $\nu$, $\Sigma_\alpha$, $\Sigma_\beta$, $\Sigma_\gamma$, and $\Sigma_\delta$ are real quantities. As a consequence, the operator eigenvalues in the transition region can be approximated as
\begin{align}
    &\lambda_q^{\op S^k} \sim \frac{\pi}{2}q^{-7/3}\left(\mu \Sigma_\alpha q^{4/3} -\nu\Sigma_\beta \right)\left( (\sqrt{3}+\mathrm{j})\nu\Sigma_\beta  + (\sqrt{3}-\mathrm{j})\mu\Sigma_\alpha q^{4/3} \right)\,,\\
    &\lambda_q^{\op N^k} \sim \frac{\pi}{2}q^{-5/3}\left(\nu\Sigma_\gamma q^{2/3} -\mu\Sigma_\delta \right)\left( (\sqrt{3}-\mathrm{j})\mu\Sigma_\delta + (\sqrt{3}+\mathrm{j})\nu\Sigma_\gamma q^{2/3} \right)\,,\\
    &\lambda_q^{\text{TM-MFIO}^k} \sim \frac{\pi}{2}q^{-2}\left(\nu \Sigma_\gamma q^{2/3} -\mu\Sigma_\delta \right)\left( (\sqrt{3}+\mathrm{j})\nu\Sigma_\beta  + (\sqrt{3}-\mathrm{j})\mu\Sigma_\alpha q^{4/3} \right)\,,\\
    &\lambda_q^{\text{TE-MFIO}^k} \sim \frac{\pi}{2}q^{-2}\left(\mu\Sigma_\alpha q^{4/3} -\nu\Sigma_\beta \right)\left( (\sqrt{3}-\mathrm{j})\mu\Sigma_\delta + (\sqrt{3}+\mathrm{j})\nu\Sigma_\gamma q^{2/3} \right)\,,\\
    &\lambda_q^{\op D^k} \sim \frac{\pi}{2}q^{-2}\bigg( 
    -\mathrm{j}\mu\nu(q^2\Sigma_\alpha\Sigma_\gamma+\Sigma_\beta\Sigma_\delta)
   + (\sqrt{3}+\mathrm{j})\nu^2 \Sigma_\beta \Sigma_\gamma q^{2/3}-
    (\sqrt{3}-\mathrm{j})\mu^2 \Sigma_\alpha \Sigma_\delta q^{4/3}
    \bigg)\,.
\end{align}
These expressions show that, in the transition region, the eigenvalues of the single-layer operator behave as $q^{1/3} \simeq (ka)^{1/3}$ in both real and imaginary parts, while the eigenvalues of the hypersingular operator behave as $q^{-1/3}\simeq (ka)^{-1/3}$ in both real and imaginary parts. In addition, the eigenvalues of the MFIO are bounded in the high-frequency limit in real and imaginary parts, consistently with the constant behavior of the imaginary part of $\lambda_q^{\op D^k}$ and with the decrease of the real part of $\lambda_q^{\op D^k}$, decaying in frequency as $q^{-2/3}\simeq (ka)^{-2/3}$.

\subsubsection{Eigenvalues in the elliptic region}
\label{sec:EAnal}
The behavior of the eigenvalues of the operators in the elliptic region \cite{buffa2006acoustic,warnick2000accuracy} can be studied by resorting to the large order asymptotic expansions of the Bessel and Hankel functions \cite[Eq. 10.19.1 and 10.19.2]{olver2010nist} and on their derivatives with respect to the argument
\begin{align}
    J_\nu^\prime(z) &\sim \sqrt{\frac{\nu}{2\pi}}\left(\frac{e z}{2\nu}\right)^\nu \frac{1}{z}\,,\\
    H_\nu^{(2)^\prime}(z) &\sim -\mathrm{j} \sqrt{\frac{2}{\nu\pi}}\left(\frac{e z}{2\nu}\right)^{-\nu} \frac{\nu}{z}\,,
\end{align}
from which we obtain
\begin{align}
    \lambda_q^{\op S^k} &\sim \frac{k a}{2 q}\,,\\
    \lambda_q^{\op N^k} &\sim \frac{q}{2 k a}\,,\\
    \lambda_q^{\text{TM-MFIE}^k} &\sim \frac{1}{2}\,,\\
    \lambda_q^{\text{TE-MFIE}^k} &\sim \frac{1}{2}\,.
\end{align}
Moreover, by expanding the derivatives of the Bessel and Hankel functions as in \cite[Eq. 10.6.1]{olver2010nist} and employing the expansions in \cite[Eq. 10.19.1 and 10.19.2]{olver2010nist}, we obtain
\begin{equation}
    \lambda_q^{\op D^k} \sim \frac{1}{16 \e}\left[4\left(\left(\frac{q-1}{q}\right)^{\frac{1}{2}-q}-\left(\frac{q+1}{q}\right)^{\frac{1}{2}+q}\right) + \e^2 (ka)^2\left(\frac{q^{-\frac{1}{2}-q}}{(q-1)^{\frac{3}{2}-q}}-\frac{q^{-\frac{1}{2}+q}}{(q+1)^{\frac{3}{2}+q}}\right)\right]\,.
\end{equation}
These expansions show a decay of the eigenvalues $\lambda_q^{\op S^k}$ and $\lambda_q^{\op D^k}$ in the elliptic region of the spectrum respectively as $q^{-1}$ and $q^{-3}$, while the eigenvalues $\lambda_q^{\op N^k}$ increase instead as $q$.
Finally, we are interested on the behavior of eigenvalues of index $q \simeq \frac{n_\lambda}{2}(ka)$, at the edge of the visible discrete bandwidth. From equations above, it follows that it is constant for the single-layer and hypersingular operators cases, while it decreases as $(ka)^{-1}$ in the double-layer operator case.

\subsection{Aliasing error of operators}
From the information about the high-frequency behavior of the eigenvalues of the operators in the three different regions, we can infer the high-frequency behavior of the aliasing spectral error of operators.

By comparing the single-layer and hypersingular operators, an asymmetry between them appears clearly: the modulus of their eigenvalues is bounded in frequency in the hyperbolic region and at the edge of the visible bandwidth, while it increases and decreases respectively as $(ka)^{1/3}$ and $(ka)^{-1/3}$ in the transition region. It follows that the aliasing spectral error respectively decreases as $(ka)^{-1/3}$ and increases as $(ka)^{1/3}$ in the transition region for the single-layer and hypersingular operators.
In the case of the single-layer operator, the spectral aliasing error is often negligible with respect to the projection error, which asymptotically dictates a bounded-in-frequency relative spectral difference between the continuous and discrete single-layer operator \eqref{eqn:spectralerror} in the transition region. In the case of the hypersingular operator, instead, both the projection or the aliasing error in the summation \eqref{eqn:spectralerror} can be dominant in the transition region depending on the discretization parameter $n_\lambda$ and on the value of $ka$. In particular, given the increase in frequency of one spectral error component, while the other stays bounded, for a fixed discretization $n_\lambda$, one may always identify the threshold frequency $(ka)_{\text{th}}$ above which the aliasing error $|E_{(ka)}^{A,\op N^k}|$ is dominant with respect to the projection error $|E_{(ka)}^P|$. Overall, the aliasing spectral error of the hypersingular operator determines an asymptotic increase of the relative spectral difference between the operator and its discrete counterpart in the transition region as $(ka)^{1/3}$ in the high-frequency regime.

Regarding the double-layer operator, given the constant behavior of the modulus of the eigenvalues in the transition region and the decay as $(ka)^{-1}$ at the bandwidth edge, the spectral aliasing error $|E_q^{A,\op D^k}|$ decreases as $(ka)^{-1}$ in the transition region. Moreover, as the imaginary part of the eigenvalues of $\op D$ is dominant in the transition region, $E_q^{A,\op D^k}$ is characterized by a different high-frequency behavior in the real and imaginary components, that is, $|\Re(E_q^{A,\op D^k})|$ behaves as $(ka)^{-5/3}$ and $|\Im(E_q^{A,\op D^k})|$ behaves as $(ka)^{-1}$.

To summarize, in the transition region in the high-frequency limit,
\begin{align}
    &|\Re(E_q^{A,\op S^k})| = \mathcal{O}((ka)^{-1/3})\,,\quad |\Im(E_q^{A,\op S^k})| = \mathcal{O}((ka)^{-1/3})\,,\label{eqn:sperrorS}\\
    &|\Re(E_q^{A,\op N^k})| = \mathcal{O}((ka)^{1/3})\,,\quad |\Im(E_q^{A,\op N^k})| = \mathcal{O}((ka)^{1/3})\,,\label{eqn:sperrorN}\\
    &|\Re(E_q^{A,\op D^k})| = \mathcal{O}((ka)^{-5/3})\,,\quad |\Im(E_q^{A,\op D^k})| = \mathcal{O}((ka)^{-1})\,. \label{eqn:sperrorD}
\end{align}

\subsection{Aliasing error of the MFIO}
Using \eqref{eqn:opsum}, the spectral relative difference between the continuous MFIO and its discretization can be written
\begin{align}
    E_q^{\text{TM-MFIO}}&=\frac{\hat{\lambda}_q^{\text{TM-MFIO}}-{\lambda}_q^{\text{TM-MFIO}}}{{\lambda}_q^{\text{TM-MFIO}}}=E_q^P + \frac{\frac{1}{2}E_q^{A,\op{I}} + \lambda_q^{\op{D}^*}E_q^{A,\op{D}^*}}{\frac{1}{2}+\lambda_q^{\op{D}^*}}\,,\\
    E_q^{\text{TE-MFIO}}&=\frac{\hat{\lambda}_q^{\text{TE-MFIO}}-{\lambda}_q^{\text{TE-MFIO}}}{{\lambda}_q^{\text{TE-MFIO}}}=E_q^P + \frac{\frac{1}{2}E_q^{A,\op{I}} - \lambda_q^{\op{D}}E_q^{A,\op{D}}}{\frac{1}{2}-\lambda_q^{\op{D}}}\,.
\end{align}
Hence, the spectral error is composed of two parts, which are, a projective component $E_q^P$, irrespective of the operators under test, and an aliasing component, given by the weighted arithmetic average between the aliasing spectral errors of the operators summed. 

As evidenced by \eqref{eqn:sperrorD}, the aliasing error contribution $|E_q^{A,\op{D}}|$ in the transition regions decays as the inverse of the frequency. However, in the high-frequency limit, this is shadowed by the aliasing contribution from the identity operator, which results in a bounded-in-frequency aliasing (and overall) spectral error of the MFIO. Overall, in the transition region in the high-frequency limit,
\begin{equation}
    |\Re(E_q^{A,\text{TM/TE-MFIO}^k})| = \mathcal{O}(1)\,,\quad |\Im(E_q^{A,\text{TM/TE-MFIO}^k})| = \mathcal{O}(1)\,.
\end{equation}

\subsection{Aliasing error of the CCFIO}
The $q-$th eigenvalue of a sum of products of commuting operators is given by the same sum of products of the $q-$th eigenvalues of these operators. Hence, the $q-$th eigenvalue of the Calderón combined field integral operator is given by
\begin{align}
    \lambda_q^{\text{TM-CCFIO}}&= \lambda_q^{\op{N}^{\tilde{k}}}\lambda_q^{\op{S}^{{k}}}+\lambda_q^{\text{TE-MFIO}^{\tilde{k}}}\lambda_q^{\text{TM-MFIO}^{{k}}}\eqcolon \lambda_q^{\text{TM-CEFIE}}+\lambda_q^{\text{TM-CMFIE}}\,,\\
    \lambda_q^{\text{TE-CCFIO}}&= \lambda_q^{\op{S}^{\tilde{k}}}\lambda_q^{\op{N}^{{k}}}+\lambda_q^{\text{TM-MFIO}^{\tilde{k}}}\lambda_q^{\text{TE-MFIO}^{{k}}}\eqcolon \lambda_q^{\text{TE-CEFIE}}+\lambda_q^{\text{TE-CMFIE}}\,.
\end{align}
We next need to estimate the relative spectral difference between the eigenvalues of the continuous operators and the ones of the discretizing matrices resulting from the matrix products and sums in the definitions of $\mat C^{\text{TM}}$ \eqref{eqn:CTM} and $\mat C^{\text{TE}}$ \eqref{eqn:CTE}.
Since all the contributions in the TM- and TE-CCFIO are commuting, their eigenvalues are respectively given by
\begin{align}
    \hat{\lambda}_q^{\text{TM-CCFIO}}&= \frac{\hat{\lambda}_q^{\op{N}^{\tilde{k}}}\hat{\lambda}_q^{\op{S}^{{k}}}}{\hat{\lambda}_q^{\op{I}}}+\frac{(\hat{\lambda}_q^{\op{I}}/2-\hat{\lambda}_q^{\op{D}^{\tilde{k}}})(\hat{\lambda}_q^{\op{I}}/2+\hat{\lambda}_q^{\op{D}^{*k}})}{\hat{\lambda}_q^{\op{I}}}\,,\\
    \hat{\lambda}_q^{\text{TE-CCFIO}}&= \frac{\hat{\lambda}_q^{\op{S}^{\tilde{k}}}\hat{\lambda}_q^{\op{N}^{{k}}}}{\hat{\lambda}_q^{\op{I}}}+\frac{(\hat{\lambda}_q^{\op{I}}/2+\hat{\lambda}_q^{\op{D}^{*\tilde{k}}})(\hat{\lambda}_q^{\op{I}}/2-\hat{\lambda}_q^{\op{D}^{{k}}})}{\hat{\lambda}_q^{\op{I}}}\,.
\end{align}
Using \eqref{eqn:opprod}, the relative spectral differences are given by the weighted means
\begin{align}
    E_q^{\text{TM-CCFIO}} = \frac{\lambda_q^{\text{TM-CEFIO}}E_q^{\text{TM-CEFIO}}+\lambda_q^{\text{TM-CMFIO}}E_q^{\text{TM-CMFIO}}}{\lambda_q^{\text{TM-CEFIO}}+\lambda_q^{\text{TM-CMFIO}}}\,,\\
    E_q^{\text{TE-CCFIO}} = \frac{\lambda_q^{\text{TE-CEFIO}}E_q^{\text{TE-CEFIO}}+\lambda_q^{\text{TE-CMFIO}}E_q^{\text{TE-CMFIO}}}{\lambda_q^{\text{TE-CEFIO}}+\lambda_q^{\text{TE-CMFIO}}}\,,
\end{align}
where the error contributions are 
\begin{align}
    E_q^{\text{TM-CEFIO}} &= \frac{(1+E_q^{\op N^{\tilde{k}}})(1+E_q^{\op S^{{k}}})}{(1+E_q^{\op I})}-1\,,\\
    E_q^{\text{TM-CMFIO}} &= \frac{(1+E_q^{\text{TE-MFIO}^{\tilde{k}}})(1+E_q^{\text{TM-MFIO}^{{k}}})}{(1+E_q^{\op I})}-1\,,\\
    E_q^{\text{TE-CEFIO}} &= \frac{(1+E_q^{\op S^{\tilde{k}}})(1+E_q^{\op N^{{k}}})}{(1+E_q^{\op I})}-1\,,\\
    E_q^{\text{TE-CMFIO}} &= \frac{(1+E_q^{\text{TM-MFIO}^{\tilde{k}}})(1+E_q^{\text{TE-MFIO}^{{k}}})}{(1+E_q^{\op I})}-1\,.
\end{align}
The weights $\lambda_q^{\text{TX-CEFIO}}$ and $\lambda_q^{\text{TX-CMFIO}}$ are of similar importance away from resonances. As the spectral error $|E_q^{\text{TX-MFIO}}|$ has a constant behavior in the high-frequency regime, the error contributions $|E_q^{\text{TM-CMFIO}}|$ and $|E_q^{\text{TE-CMFIO}}|$ will also be characterized by a constant behavior. Differently, $|E_q^{\text{TM-CEFIO}}|$ and $|E_q^{\text{TE-CEFIO}}|$ will be negatively affected by the increase of $|E_q^{\op N}|$. In conclusion, this will result in an asymptotically increasing behavior of $|E_q^{\text{TM-CCFIO}}|$ and $|E_q^{\text{TE-CCFIO}}|$ as $(ka)^{1/3}$ in the transition region,
\begin{align}
    &|\Re(E_q^{A,\text{TM-CCFIO}})| = \mathcal{O}((ka)^{1/3})\,,\quad |\Im(E_q^{A,\text{TM-CCFIO}})| = \mathcal{O}((ka)^{1/3})\,    \\
    &|\Re(E_q^{A,\text{TE-CCFIO}})| = \mathcal{O}((ka)^{1/3})\,,\quad |\Im(E_q^{A,\text{TE-CCFIO}})| = \mathcal{O}((ka)^{1/3})\,.
\end{align}

\section{Current error analysis}
\label{sec:currenterror}
This section aims at studying the consequences of the high-frequency behaviors of the aliasing spectral errors presented in Section~\ref{sec:spectralerror} on the current errors, defined as the relative difference between the solutions of the integral equations \Cref{eqn:TMEFIE},\eqref{eqn:TMMFIE},\eqref{eqn:TEEFIE},\eqref{eqn:TEMFIE},\eqref{eqn:TMCCFIE},\eqref{eqn:TECCFIE} and currents resulting from the solution of the BEM systems \eqref{eqn:TMEFIEd},\eqref{eqn:TMMFIEd},\eqref{eqn:TEEFIEd},\eqref{eqn:TEMFIEd},\eqref{eqn:CTM},\eqref{eqn:CTE}. 
These two currents will be denoted respectively as $J$ and $\hat{J}$.
The excitation considered is the plane wave incident to the cylinder \cite{warnick2000accuracy,warnick2008numerical}.

From the Mie scattering solution from a PEC circular cylinder, the value $J_n$ of the current $J$ evaluated at the point with polar coordinates $(a,\phi_n)$ can be written as \cite{warnick2000accuracy,warnick2008numerical}
\begin{equation}
    J_n = \sum_{q=-\infty}^\infty U_q \, e^{-\mathrm{j}q\phi_n}
\end{equation}
where $U_q$ is given by
\begin{align}
    U_q^{\text{TM}} &= \frac{2 \mathrm{j}^{-q}}{\pi \eta k a \, H_q^{(2)}(k a)} \,,\label{eqn:UTM}\\
    U_q^{\text{TE}} &= \frac{2 \mathrm{j}^{-q}}{\pi \eta k a \, H_q^{(2)^\prime}(k a)}\label{eqn:UTE}\,.
\end{align}
Similarly, the current resulting from the solution of one of the BEM linear systems evaluated at the point $(a,\phi_n)$, corresponding to the $n-$th element of the array $\hat{\vec J}$, is denoted as $\hat{J}_n$ and can be written as
\begin{equation}
    \hat{J}_n = \sum_{q=-\infty}^\infty \hat{U}_q \, e^{-\mathrm{j}q\phi_n}\,.
\end{equation}
The difference between the solution of the integral equation and the solution of the linear system is denoted as $\Delta J$. The value $\Delta J_n$ of $\Delta J$ at $(a,\phi_n)$ is the difference between $\hat{J}_n$, representing the solution of the discrete problem, and $J_n$, representing the solution of the continuous problem at the point $(a,\phi_n)$. 
After defining the relative error coefficient $\upsilon_q$ as $\upsilon_q \coloneqq (\hat{U}_q-{U}_q)/U_q$, we expand the difference as
\begin{equation}
    \Delta J_n = \hat{J}_n- J_n = \sum_{q=-\infty}^\infty U_q \upsilon_q \, e^{-\mathrm{j}q\phi_n}\,.
\end{equation}

For the EFIE, the $\upsilon_q$ coefficient reads \cite{warnick2000accuracy,warnick2008numerical}
\begin{equation} 
    \upsilon_q^{\text{TM/TE-EFIE}}=\frac{F_{-q}(1-F_q)-E_q^{A,\op{S}^k/\op{N}^k}}{1+E_q^{\op{S}^k/\op{N}^k}}\,.
    \label{eqn:upsilonEFIE}
\end{equation}
By following similar steps as in \cite{warnick2008numerical}, that is, by expressing the solution of the discrete problem as the ratio between the Fourier mode expansions of the right- and left-hand sides (RHS and LHS), for the MFIE we have
\begin{equation} 
    \upsilon_q^{\text{TM/TE-MFIE}}=\frac{F_{-q}(1-F_q)-E_q^{A,\text{TM/TE-MFIO}^k}}{1+E_q^{\text{TM/TE-MFIO}^k}}\,.
    \label{eqn:upsilonMFIE}
\end{equation}

To determine the current error in the CCFIE case, we sum the Fourier coefficients of the two components of the RHS tested with pyramid basis functions and divide the result by the Fourier coefficients of the discretization of the operator. The derivation, detailed in \ref{sec:appendix}, leads to the definition of $\upsilon_q^{\text{CCFIE}}$ as
\begin{equation}
\upsilon_q^{\text{TM/TE-CCFIE}}=\frac{\hat{\lambda}_q^{\text{TM/TE-CEFIO}}\upsilon_q^{\text{TM/TE-EFIE}}+\hat{\lambda}_q^{\text{TM/TE-CMFIO}}\upsilon_q^{\text{TM/TE-MFIE}}}{\hat{\lambda}_q^{\text{TM/TE-CEFIO}}+\hat{\lambda}_q^{\text{TM/TE-CMFIO}}}\,.
\end{equation}

\subsection{Current error measures}
Different measures of the current error are possible. We propose here some of the most commonly used, useful and significant for diverse purposes.

We consider here the $L^2$-measure,
\begin{equation}
    r_{L^2(\Gamma)} \coloneqq \left(\frac{\sum_{q=-\infty}^\infty|U_q \upsilon_q|^2}{\sum_{q=-\infty}^\infty|U_q|^2} \right)^{1/2}\,,
    \label{eqn:rl2}
\end{equation}
the measure in the standard norm of the current space $H^s(\Gamma)$, with $s=\mp 1/2$ for the TM/TE formulations,
\begin{equation}
    r_{H^s(\Gamma)} \coloneqq\left(\frac{\sum_{q=-\infty}^\infty|U_q \upsilon_q|^2\, (1+q^2)^s}{\sum_{q=-\infty}^\infty|U_q|^2\, (1+q^2)^s} \right)^{1/2}\,,
    \label{eqn:rH}
\end{equation}
and the measure in a different norm in $H^s(\Gamma)$,
\begin{equation}
    r_{H_k^s(\Gamma)} \coloneqq\left(\frac{\sum_{q=-\infty}^\infty|U_q \upsilon_q|^2\, (k^2+q^2)^s}{\sum_{q=-\infty}^\infty|U_q|^2\, (k^2+q^2)^s} \right)^{1/2}\,,
    \label{eqn:rHk}
\end{equation}
commonly used in high-frequency scattering applications \cite{chandler-wilde2015wavenumberexplicit,galkowski2022does}. Another common error measure is the one in the $|\cdot|_{P^{\text{TM/TE}}}$ semi-norm \cite{davis2005physical}
\begin{equation}
    r_{P^{\text{TM/TE}}(\Gamma)} \coloneqq\left|\frac{\sum_{q=-\infty}^\infty|U_q \upsilon_q|^2\, \alpha_q^{\text{TM/TE}}}{\sum_{q=-\infty}^\infty|U_q|^2\, \alpha_q^{\text{TM/TE}}} \right|^{1/2}\,,
    \label{eqn:pnorm}
\end{equation}
where
\begin{align}
    \alpha_q^{\text{TM}} &=\frac{k\eta \pi a}{2} J_q(ka) H_q^{(2)}(ka)\,,\label{eqn:alphaTM}\\
    \alpha_q^{\text{TE}} &=\frac{k\eta \pi a}{2} J_q^\prime(ka) H_q^{(2)\prime}(ka)\,.\label{eqn:alphaTE}
\end{align}
So, given the expressions \eqref{eqn:alphaTM}, \eqref{eqn:alphaTE} and the eigenvalues of the single-layer and hypersingular operators \eqref{eqn:lamS}, \eqref{eqn:lamN}, equation \eqref{eqn:pnorm} corresponds to
\begin{equation}
    r_{P^{\text{TM}}(\Gamma)}= \left|\frac{\left( \op S^k  \Delta J,  \Delta J\right)_{L^2(\Gamma)}}{\left( \op S^k J, J\right)_{L^2(\Gamma)}} \right|^{1/2}\quad \text{and}\quad
    r_{P^{\text{TE}}(\Gamma)}= \left|\frac{\left( \op N^k  \Delta J,  \Delta J\right)_{L^2(\Gamma)}}{\left( \op N^k J, J\right)_{L^2(\Gamma)}} \right|^{1/2}
\end{equation}
which can be approximated on any geometry by substituting the continuous operators with their discrete counterparts numerically available,
\begin{equation}
    \tilde{r}_{P^{\text{TM}}(\Gamma)}\coloneqq \left|\frac{(\hat{\vec J}-\vec J)^\T\mat S^k  (\hat{\vec J}-\vec J)}{\vec J^\T \mat S^k \vec J} \right|^{1/2}\simeq {r}_{P^{\text{TM}}(\Gamma)}\quad \text{and}\quad
    \tilde{r}_{P^{\text{TE}}(\Gamma)}\coloneqq \left|\frac{(\hat{\vec J}-\vec J)^\T\mat N^k  (\hat{\vec J}-\vec J)}{\vec J^\T \mat N^k \vec J} \right|^{1/2}\simeq {r}_{P^{\text{TE}}(\Gamma)}\,,
\end{equation}
where $\vec J$ is the array of coefficients of the expansion of the current $J$ in the pyramid functional space.

As a general remark, we notice that, differently from the $H^s_k(\Gamma)$ norm \cite{chandler-wilde2015wavenumberexplicit}, the $|\cdot|_{P^{\text{TM/TE}}}$ semi-norm does not preserve the balance between lower and higher derivatives. The higher derivatives of the current, corresponding to spectral coefficients in the transition region, are scaled by factors which are $\mathcal{O}(ka)^{1/3}$ or $\mathcal{O}(ka)^{-1/3}$ for the TM/TE formulations respectively. Hence, their relative importance with respect to lower derivatives, corresponding to spectral indices in the hyperbolic region, increases/decreases in frequency as $(ka)^{\pm 1/3}$.
In addition, while the $P^{\text{TM/TE}}$ semi-norm is equivalent to the $H^s$ and to the $H^s_k$ norms with respect to the discretization parameter $h$, it is not equivalent to them with respect to the frequency \cite{chandler-wilde2015wavenumberexplicit,davis2005physical}, i.e.,
\begin{equation}
   c_1(k)\,|u|_{P^{\text{TM/TE}}}\le \|u\|_{H^{\mp 1/2}}\le c_2(k)\,|u|_{P^{\text{TM/TE}}}\,,
\end{equation}
where the parameters $c_1$, $c_2$ are not bounded in frequency. For these reasons, the $P^{\text{TM/TE}}$ semi-norm will not be considered further in this work.

\subsection{High-frequency analysis of the current error}
\label{sec:erroranalysisHF}
The coefficient $|U_q|^2$ in equations \eqref{eqn:rl2}, \eqref{eqn:rH}, and \eqref{eqn:rHk} is a non-oscillating function of the spectral index $q$. $|U_q^{\text{TM}}|^2$ scales as $(ka)^{-1}$ in the hyperbolic region and as $(ka)^{-4/3}$ in the transition region; conversely, $|U_q^{\text{TE}}|^2$ scales as $(ka)^{-1}$ in the hyperbolic region and as $(ka)^{-2/3}$ in the transition region.
Moreover, it decays to zero in the elliptic region, making the contributions of the summations corresponding to indices $|q| \gg ka$ negligible, so that we safely restrict our analysis to the contributions in the hyperbolic and transition regions.

From the spectral analyses above, we infer that $|\upsilon_q^{\text{TM-EFIE}}|$ and $|\upsilon_q^{\text{TM/TE-MFIE}}|$ have a constant behavior in the high-frequency limit, both in the hyperbolic and transition region. As a consequence, the current relative error of the corresponding formulations is expected to be bounded in the high-frequency regime. 

Differently, the increase of $|E_q^{A,\op N^k}|$ as $(ka)^{1/3}$ in the high-frequency limit leads to an increase of $|\upsilon_q^{\text{TE-EFIE}}|$ in the transition region at the same rate. The absence of lossy terms in the electric field equation (i.e., the kernel involved is the lossless Green's function $G^k$) results in a spectral shape invariance in frequency, so that the number of eigenvalues of $\op N^k$ decaying in modulo as $(ka)^{-1/3}$ is proportional to the frequency. 
Hence, in \eqref{eqn:rl2}, $\sum_{q=-\infty}^\infty|U_q \upsilon_q|^2$ can be seen as the sum of $\mathcal{O}(ka)$ terms in the transition region behaving as $(ka)^{-2/3}\cdot(ka)^{2/3}$ and of $\mathcal{O}(ka)$ terms in the hyperbolic region behaving as $(ka)^{-1}\cdot(ka)^{0}$, from which $\sum_{q=-\infty}^\infty|U_q \upsilon_q|^2=\mathcal{O}(ka)$. Similarly for the denominator in \eqref{eqn:rl2}, $\sum_{q=-\infty}^\infty|U_q|^2$, which approximately corresponds to the summation of $\mathcal{O}(ka)$ terms in the transition region behaving as $(ka)^{-2/3}$ and of $\mathcal{O}(ka)$ terms in the hyperbolic region behaving as $(ka)^{-1}$, from which $\sum_{q=-\infty}^\infty|U_q|^2=\mathcal{O}((ka)^{1/3})$. Finally, when considering the TE-EFIE, the error measure $r_{L^2(\Gamma)}$ increases asymptotically as $(ka)^{1/3}$ in the high-frequency limit. By following a similar reasoning, the high-frequency behavior as $(ka)^{1/3}$ of $r_{H^s(\Gamma)}$ and $r_{H^s_k(\Gamma)}$ for the TE-EFIE can be shown.

In the TM-CCFIE case, the error contribution $\upsilon_q^{\text{TM-CCFIE}}$ is given by the weighted average between the two contributions $\upsilon_q^{\text{TM-EFIE}}$ and $\upsilon_q^{\text{TM-MFIE}}$, which are bounded in the high-frequency limit. Hence, in this case, the increasing spectral error does not result in an increasing current error, as the increasing behavior of $|E_q^{A,\op N^{\tilde{k}}}|$ in frequency only impacts the relative importance of $\upsilon_q^{\text{TM-EFIE}}$ and $\upsilon_q^{\text{TM-MFIE}}$, one with respect to each other: the spectral deformation induced by the preconditioning does not induce any deterioration in the current accuracy.
Moreover, we note that, even though the two error components $\upsilon_q^{\text{TM-EFIE}}$ and $\upsilon_q^{\text{TM-MFIE}}$ may present sharp peaks in correspondence of determined indices $q_{\text{res,TM-EFIE}}$ and $q_{\text{res,TM-MFIE}}$ at the resonance frequencies of the TM-EFIE and TM-MFIE, the current error of the TM-CCFIE is immune from spurious resonances as a result of the weighted average in favor of $\upsilon_q^{\text{TM-MFIE}}$ at indices $q_{\text{res,TM-EFIE}}$ and in favor of $\upsilon_q^{\text{TM-EFIE}}$ at indices $q_{\text{res,TM-MFIE}}$.

The analysis is more complicated in the TE-CCFIE case. At first, we notice that the presence of lossy contributions in the equation (i.e., the presence of the complex $\tilde{k}$ in the kernel of the preconditioning operators) determines a spectral deformation in frequency. For spectral indices in the transition region, the weighted average with oscillating weights alleviates the effect of the high-frequency increase of $|\upsilon_q^{\text{TE-EFIE}}|$ over $|\upsilon_q^{\text{TE-CCFIE}}|$. Hence, we expect the current error of the TE-CCFIE to increase in the high-frequency limit at a rate equal or slower than $(ka)^{1/3}$. Numerical results in \Cref{sec:numresults} indicate a substantial improvement of the current error increase rate of the TE-CCFIE with respect to the TE-EFIE.

\section{Scattering Error Analysis}
\label{sec:scatteringerror}
Given a plane wave traveling along a transverse direction and impinging on the cylindrical scatterer, the electric (for TM polarization) or magnetic (for TE polarization) scattered field in the longitudinal direction is proportional to the scattering parameter \cite{warnick2008numerical} 
\begin{equation}
    S(\phi) = \sum_{q=-\infty}^\infty R_q e^{-\mathrm{j} q \phi}\,,
\end{equation}
where the expression of the Fourier coefficients $R_q$ for the two polarizations is
\begin{align}
    R_q^{\text{TM}} &= J_q(ka) / H_q^{(2)}(ka)\,,\label{eqn:RTM}\\
    R_q^{\text{TE}} &= J_q^\prime(ka) / H_q^{(2)^\prime}(ka)\label{eqn:RTE}\,.
\end{align}
In particular, for the TM polarization, the electric far field scattered by the longitudinal current $\veg J_z$ at the point at polar coordinates $(r_{\text{FF}},\phi)$ for $r_{\text{FF}} \gg a$ is
\begin{equation}
    \veg E(r_{\text{FF}},\phi) \simeq -\sqrt{\frac{2\mathrm{j}}{\pi k r_{\text{FF}}}} e^{-\mathrm{j}kr_{\text{FF}}} S^{\text{TM}}(\phi) \hat{\veg{z}}\,,
\end{equation}
where the far field approximation \cite[Eq. 2.3.17]{jin2015theory}
and the large argument expansion of the Hankel function \cite[Eq. 9.2.4]{abramowitz1964handbook} have been applied.
Similarly for the TE polarization, by applying equivalent approximations, the far field magnetic field reads 
\begin{equation}
    \veg H(r_{\text{FF}},\phi) \simeq -\frac{1}{\mathrm{j}\eta}\sqrt{\frac{2\mathrm{j}}{\pi k r_{\text{FF}}}} e^{-\mathrm{j}kr_{\text{FF}}} S^{\text{TE}}(\phi) \hat{\veg{z}}\,.
\end{equation}

On the other hand, the far field electric and magnetic fields resulting from the numerical solution of the discrete TM and TE formulations can be written as \cite{warnick2008numerical}
\begin{align}
    \hat{\veg E}(r_{\text{FF}},\phi) &\simeq -\sqrt{\frac{2\mathrm{j}}{\pi k r_{\text{FF}}}} e^{-\mathrm{j}kr_{\text{FF}}} \hat{S}^{\text{TM}}(\phi) \hat{\veg{z}}\,,\\
    \hat{\veg H}(r_{\text{FF}},\phi) &\simeq -\frac{1}{\mathrm{j}\eta}\sqrt{\frac{2\mathrm{j}}{\pi k r_{\text{FF}}}} e^{-\mathrm{j}kr_{\text{FF}}} \hat{S}^{\text{TE}}(\phi) \hat{\veg{z}}\,,
\end{align}
where
\begin{equation}
    \hat{S}(\phi) \coloneqq \sum_{q=-\infty}^\infty \hat{R}_q e^{-\mathrm{j} q \phi}\,.
\end{equation}

After defining the relative error coefficient $\rho_q$ as $\rho_q \coloneqq (\hat{R}_q-{R}_q)/R_q$, we expand the difference between the numerical scattering resulting from the solution of the discrete equations and the analytic one as
\begin{equation}
    \hat{S}(\phi) - S(\phi) = \sum_{q=-\infty}^\infty R_q \rho_q e^{-\mathrm{j} q \phi}\,.
\end{equation}
From the Fourier mode expansion of the scattering operator, the error coefficient is given by
\begin{equation}
    \rho_q = F_q (\upsilon_q+1)-1 \label{eqn:relrhoupsilon}\,.
\end{equation}
Hence, for the EFIE, the $\rho_q$ coefficient reads \cite{warnick2008numerical}
\begin{equation}
    \rho_q^{\text{TM/TE-EFIE}}=-\frac{E_q^{A,\op{S}^k/\op{N}^k}}{1+E_q^{\op{S}^k/\op{N}^k}}\,,\label{eqn:rhoEFIE}
\end{equation}
and, for the MFIE, it reads
\begin{equation}
    \rho_q^{\text{TM/TE-MFIE}}=-\frac{E_q^{A,\text{TM/TE-MFIO}^k}}{1+E_q^{\text{TM/TE-MFIO}^k}}\,.
\end{equation}

The CCFIE case can be derived from the definition of $\upsilon_q^{\text{CCFIE}}$ and \eqref{eqn:relrhoupsilon}, resulting in the expression
\begin{equation}
    \rho_q^{\text{TM/TE-CCFIE}}=\frac{\hat{\lambda}_q^{\text{TM/TE-CEFIO}}\rho_q^{\text{TM/TE-EFIE}}+\hat{\lambda}_q^{\text{TM/TE-CMFIO}}\rho_q^{\text{TM/TE-MFIE}}}{\hat{\lambda}_q^{\text{TM/TE-CEFIO}}+\hat{\lambda}_q^{\text{TM/TE-CMFIO}}}\,.
\end{equation}

\subsection{Scattering error measure}
We define the scattering relative error as
\begin{equation}
    s_{L^2(\Gamma_{\text{FF}})}\coloneqq 
    \left(\frac{\sum_{q=-\infty}^\infty|R_q \rho_q|^2}{\sum_{q=-\infty}^\infty|R_q|^2}\right)^{1/2}\,.
\end{equation}
Note that, given the definition of the coefficient $R_q$ (\eqref{eqn:RTM}, \eqref{eqn:RTE}) and the one of the coefficient $U_q$ (\eqref{eqn:UTM}, \eqref{eqn:UTE}), $s_{L^2}$ can also be written, for the two polarizations, as
\begin{align}
    s_{L^2(\Gamma_{\text{FF}})}^{\text{TM}} &\coloneqq 
    \left(\frac{\sum_{q=-\infty}^\infty|U_q^{\text{TM}} J_q(ka) \rho_q|^2}{\sum_{q=-\infty}^\infty|U_q^{\text{TM}} J_q(ka)|^2}\right)^{1/2}\,,
    \label{eqn:sTM}
    \\
    s_{L^2(\Gamma_{\text{FF}})}^{\text{TE}} &\coloneqq 
    \left(\frac{\sum_{q=-\infty}^\infty|U_q^{\text{TE}} J_q^\prime(ka) \rho_q|^2}{\sum_{q=-\infty}^\infty|U_q^{\text{TE}} J_q^\prime(ka)|^2}\right)^{1/2}\,.
    \label{eqn:sTE}
\end{align}

\subsection{High-frequency analysis of the scattering error}
\label{sec:scaterroranalysisHF}
By following the same reasoning as in Section~\ref{sec:currenterror}, given the high-frequency behavior of the aliasing spectral error of the operators in the transition region, the scattering error is expected to behave as $\mathcal{O}(1)$ for the TM-EFIE, TM-MFIE, and TE-MFIE, and as $\mathcal{O}((ka)^{1/3})$ for the TE-EFIE in the high-frequency limit. Moreover, as the scattering error of the TM-CCFIE depends on the weighted average between $\rho^{\text{TM-EFIE}}_q$ and $\rho^{\text{TM-MFIE}}_q$, bounded in the high-frequency limit, it is expected to be bounded in the high-frequency regime, similarly as the current error.
On the other hand, an increase of the scattering error at a rate at most equal to $(ka)^{1/3}$ is expected from the solution of the TE-CCFIE due to the increasing behavior of $\rho^{\text{TE-EFIE}}_q$ as $(ka)^{1/3}$ in the transition region in the high-frequency limit.

As an aside, it is worth noting that the presence of the coefficients $|J_q(ka)|^2$ and $|J_q^\prime(ka)|^2$ in the definition of $s_{L^2(\Gamma_{\text{FF}})}^{\text{TM}}$ \eqref{eqn:sTM} and $s_{L^2(\Gamma_{\text{FF}})}^{\text{TE}}$ \eqref{eqn:sTE} affects the sensitivity of the scattering error to spurious resonances differently for different formulations. Indeed, since the zeros of $J_q(ka)$ and $J_q^\prime(ka)$ determine the resonances of the TM-EFIE and TE-EFIE respectively, we expect that the coefficient $|\rho_q|^2$ at the resonant indices $q_{\text{res,TM-EFIE}}$ and $q_{\text{res,TE-EFIE}}$ will be attenuated by $|J_q(ka)|^2$ and $|J_q^\prime(ka)|^2$, approaching the zero at the same indices. This results, when solving the EFIE, in a scattering error that is quasi-immune to spurious resonances, as reported in the literature \cite{antoine2016integral,christiansen2003discrete,boeykens2013rigorous}.
The contributions $|J_q(ka)|^2$ and $|J_q^\prime(ka)|^2$, beneficial for the scattering from the EFIE, become detrimental in the MFIE case instead. 
Indeed, the resonant indices $q_{\text{res,TM-MFIE}}$ and $q_{\text{res,TE-MFIE}}$, at which the coefficients $|\rho_q|^2$ are expected to reach local maxima, are determined by the zeros of $J_q^\prime(ka)$ and $J_q(ka)$ respectively. Given the oscillatory nature of the Bessel function $J_q(ka)$, at these resonance indices, the functions $J_q(ka)$ and $J_q^\prime(ka)$ respectively achieve their local maxima, leading to an amplification of the detrimental effect of spurious resonances over the scattering error from the TM- and TE-MFIE.

\section{Overcoming the Pollution Effect by Filtering}
\label{sec:filtering}
By restricting our analysis to the cylindrical case, we loose in generality, but we gain a powerful model to explain the mechanism of discretization of integral operators by using BEM and its consequences. As the aliasing spectral error of the hypersingular operator is responsible for the increasing behavior of the current and scattering errors delineated in the above sections, we believe that spectral operator filtering strategies, such as the ones presented in \cite{masciocchi2023operator}, may significantly attenuate, or even solve, the pollution problem.

To showcase this idea, we will consider the ideally filtered operator $\op N^k_F$ with eigenvalues
\begin{equation}
    \lambda_q^{\op N_F^k} = \begin{cases}
        \lambda_q^{\op N^k}\,, & \text{if $|q|\le q_{\text{lim}}$}\\
            0\,, & \text{otherwise}
    \end{cases}\,.
\end{equation}
The choice of the cutoff spectral index
\begin{equation}
    q_{\text{lim}} = \lfloor (n_{\lambda}-1-\epsilon)ka \rfloor
\end{equation}
with $\epsilon > 0$, allows the suppression of the aliasing spectral error in both the hyperbolic and the transition regions of the spectra, as is clear from equation \eqref{eqn:aliaserror}. Hence, in the transition region in the high-frequency limit, we obtain
\begin{equation}
    |\Re(E_q^{\op N_F^k})| = \mathcal{O}(1)\,,\quad |\Im(E_q^{\op N_F^k})| = \mathcal{O}(1)\,,
\end{equation}
which in turn results in a bounded spectral error of the operators $\text{TM-CCFIO}_F$ and $\text{TE-CCFIO}_F$, with expressions given by the ones of the correspective non filtered ones where $\op N^k$ is substituted by $\op N^k_F$ and $\op N^{\tilde{k}}$ is substituted by $\op N^{\tilde{k}}_F$,
\begin{align}
    &|\Re(E_q^{\text{TM-CCFIE}_F^k})| = \mathcal{O}(1)\,,\quad |\Im(E_q^{\text{TM-CCFIE}_F^k})| = \mathcal{O}(1)\,,\\
    &|\Re(E_q^{\text{TE-CCFIE}_F^k})| = \mathcal{O}(1)\,,\quad |\Im(E_q^{\text{TE-CCFIE}_F^k})| = \mathcal{O}(1)\,.
\end{align}

The bounded-in-frequency spectral error of the filtered formulations leads in turn to a bounded current and scattering error.
To estimate the practical impact of filtering, we evaluate the current and spectral error resulting from the solution of the $\text{TE-EFIE}_F$ in which we have substituted the operator $\op N^k$ with $\op N_F^k$, and of the $\text{TE-CCFIE}_F$. The results of this analysis are reported in \Cref{sec:numresults}.

\section{Numerical Results}
\label{sec:numresults}
The numerical results presented in this section are intended to demonstrate the convergence between the numerical results from our BEM implementation and what is predicted from theory, as well as verifying numerically the high-frequency behavior of the spectral, current, and scattering errors. The BEM matrices and RHSs have been obtained (in their non-singular terms) by numerical integration exploiting standard Gauss-Legendre rules with a high number of integration points per mesh element (100), to minimize the integration error with respect to the discretization error which is the source of numerical error of interest in this contribution. The results predicted from our developments have been obtained by truncating the infinite sums in $q$ to $q\in\left[-(N-1)/2,(N-1)/2\right]$ (with $N$ odd) and by only considering the first harmonic term in the aliasing contribution, that is, limiting the infinite summation in \eqref{eqn:WarnickFormula} to the terms corresponding to $s=\{-1,0,1\}$. The only exception to this setting is in Figure~\ref{fig:sp_error_reim} in which two harmonics were taken into account, corresponding to the terms $s=\{-2,-1,0,1,2\}$, in order to improve the convergence to numerical results. In general, a further improvement of convergence has been observed and is expected for higher numbers of harmonics, but numerical stability in the evaluation of special functions limits the analysis.

If not otherwise specified, the circular geometry under test is the one of radius \SI{1}{\meter} discretized at $n_\lambda=4$. Given the circulant structure of the interaction matrices, the fast Fourier transform (FFT) algebra is employed to compute matrix-vector and diagonalization operations in linear complexity.

\subsection{Spectral error results}

\begin{figure}
\centerline{\includegraphics[width=0.5\columnwidth]{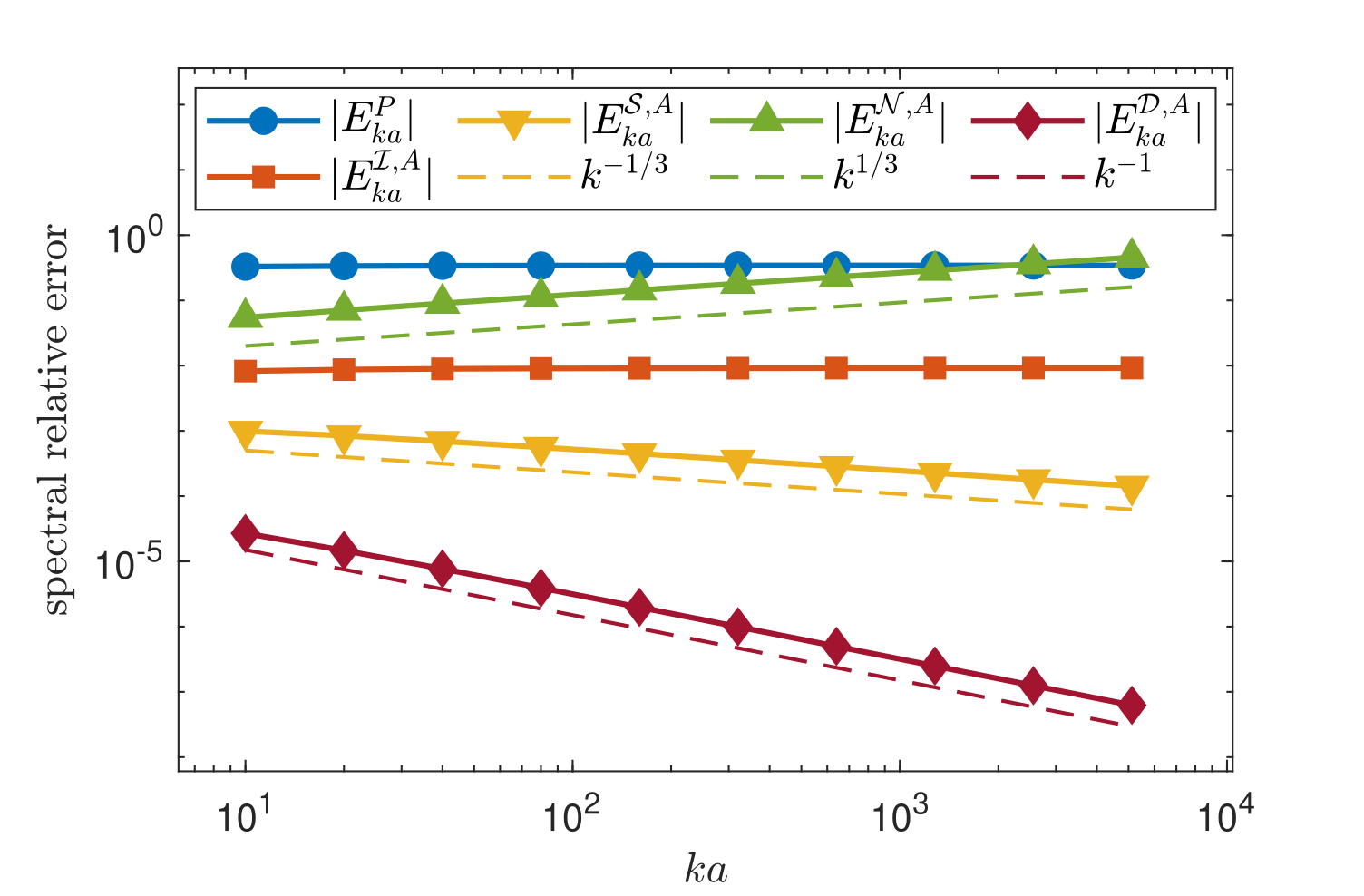}}
\caption{Predicted values (from formulae \eqref{eqn:aliaserror}) of projection and aliasing spectral error in the transition region for diverse operators for varying $ka$ with $n_\lambda=4$.}
\label{fig:sp_error_baseop}
\end{figure}

In the first set of numerical results, we analyze the high-frequency behavior of the spectral relative error of the four standard operators \eqref{eqn:aliaserror}, to confirm that the trends predicted from the asymptotic expansions of the Bessel functions in Section~\ref{sec:spectralerror} hold. \Cref{fig:sp_error_baseop} shows the predicted projection and aliasing contributions in absolute value for the operators $\mathcal{I}$, $\mathcal{S}$, $\mathcal{N}$, and $\mathcal{D}$, in the transition region, for $q=ka$ (with $ka$ integer), obtained by evaluation of \eqref{eqn:aliaserror}. The results confirm the predicted decrease as $(ka)^{-1/3}$ and as $(ka)^{-1}$ of $|E_{ka}^{A,\mathcal{S}}|$ and $|E_{ka}^{A,\mathcal{D}}|$ and the expected increase as $(ka)^{1/3}$ of $|E_{ka}^{A,\mathcal{N}}|$.

\begin{figure}
\centerline{\includegraphics[trim= 0 0 0 80,clip,width=0.8\columnwidth]{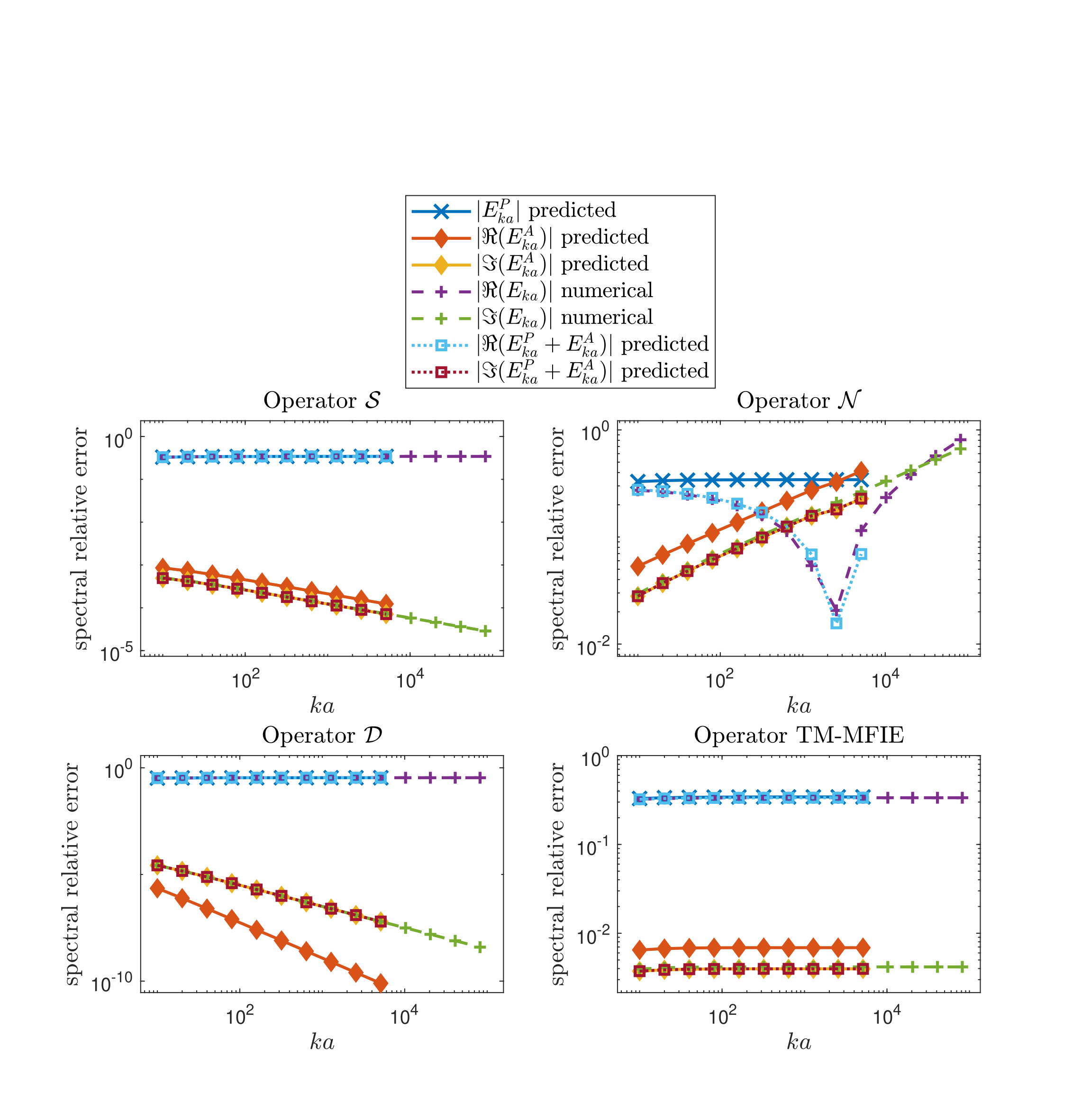}}
\caption{Projection and aliasing spectral error in the transition region for diverse operators for varying $ka$ with $n_\lambda=4$. Comparison between the predicted values (from formulae \eqref{eqn:aliaserror}) and numerical values from our BEM code.}
\label{fig:sp_error_reim}
\end{figure}

In \Cref{fig:sp_error_reim} the analytically predicted and numerical spectral relative error are compared---in real and imaginary components. We observe that the real part of $E_{ka}$ obtained numerically corresponds in modulus to $|E_{ka}^P|$ (where $E_{ka}^P$ is real) for $\mathcal{S}$, $\mathcal{D}$, and TM-MFIE operators. Instead, the imaginary part of $E_{ka}$ corresponds to the imaginary part of the aliasing error contribution and decays in modulo as $(ka)^{-1/3}$ for $\mathcal{S}$, as $(ka)^{-1}$ for $\mathcal{D}$, and is bounded for the TM-MFIO. We also observe that the real part of the aliasing error contribution for operator $\mathcal{D}$ decays in modulo as $(ka)^{-5/3}$, following from the behavior of $|\Im(\lambda_q^{\mathcal{D}})|$ in the transition region as $(ka)^{-2/3}$. In the hypersingular operator case, the real part of the aliasing component contributes to the definition of the real spectral relative error. We observe that it is of opposite sign with respect to the projection component, resulting in an initial decrease of the sum, up to a cancellation in the frequency point at which $E_{ka}^P\simeq-\Re(E_{ka}^{A,\mathcal{N}})$. However, the asymptotic behavior of $E_{ka}^{\mathcal{N}}$ in both real and imaginary parts is determined by the aliasing error component which increases in modulo as $(ka)^{1/3}$ in both real and imaginary components.

\begin{figure}
\centerline{\includegraphics[width=0.5\columnwidth]{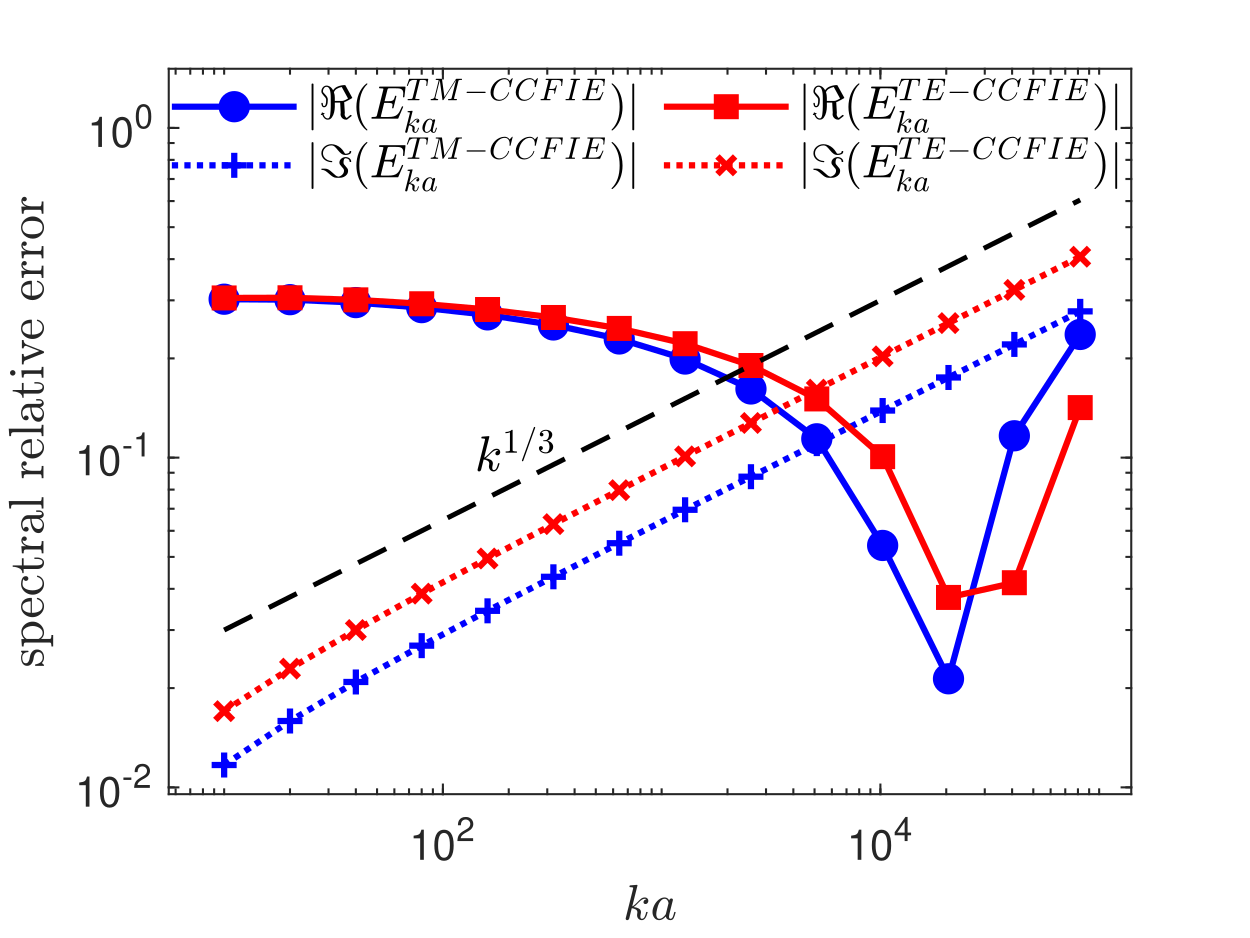}}
\caption{Spectral error of the CCFIO in the transition region for varying $ka$ with $n_\lambda=4$. These results have been obtained numerically from our BEM code.}
\label{fig:sp_error_CCFIE}
\end{figure}

The numerically obtained spectral relative error for the CCFIE operators is represented in \Cref{fig:sp_error_CCFIE}, in real and imaginary components, and shows a behavior similar to the $\mathcal{N}$ operator: the imaginary component increases as $(ka)^{1/3}$, while the real part is subject to cancellation before increasing.

\subsection{Current error results}

\begin{figure}
\centering
\subfloat[\label{subfig-1:}]{%
  \includegraphics[width=0.49
\columnwidth]{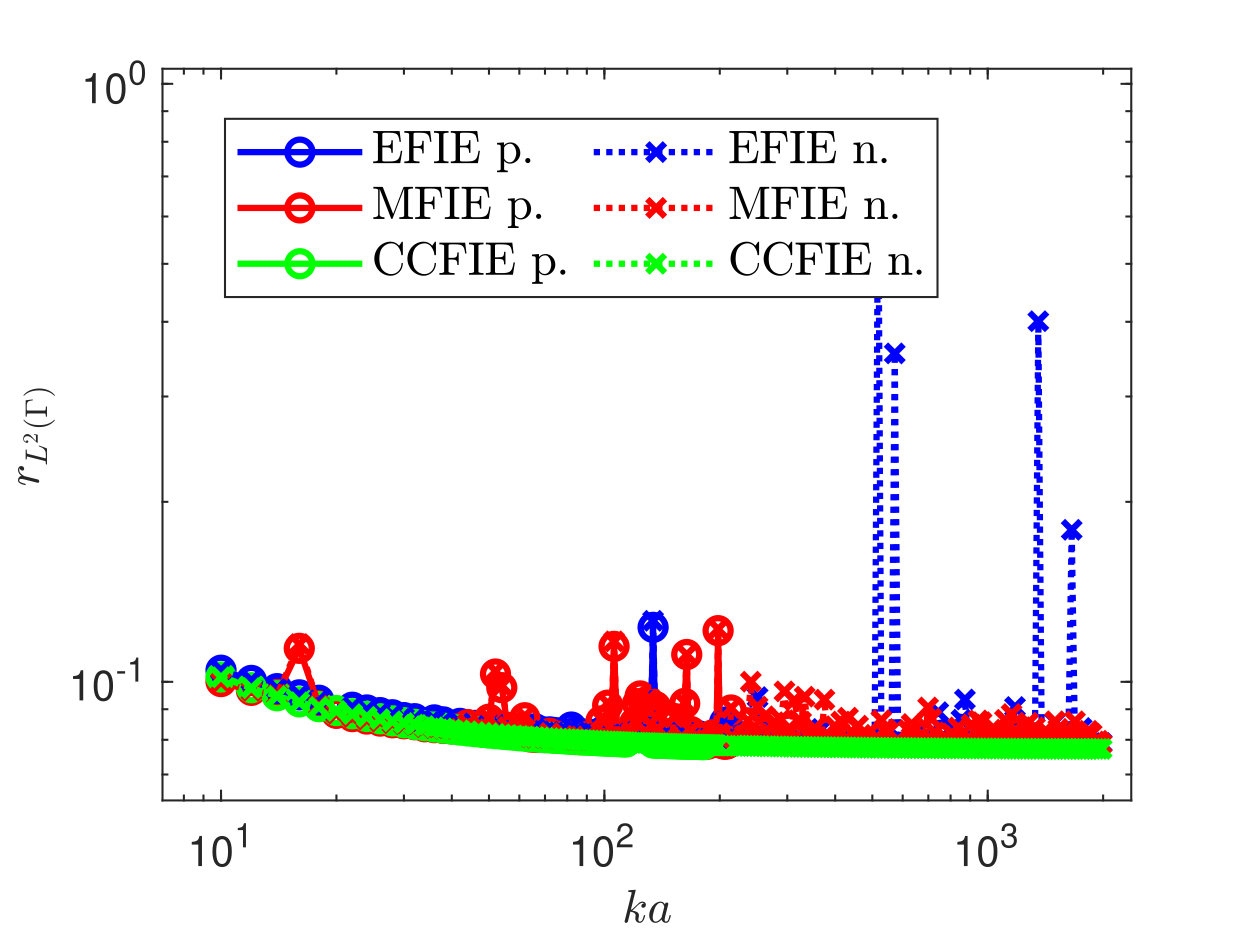}
}
\hfill
\subfloat[\label{subfig-2:}]{%
\includegraphics[width=0.49\columnwidth]{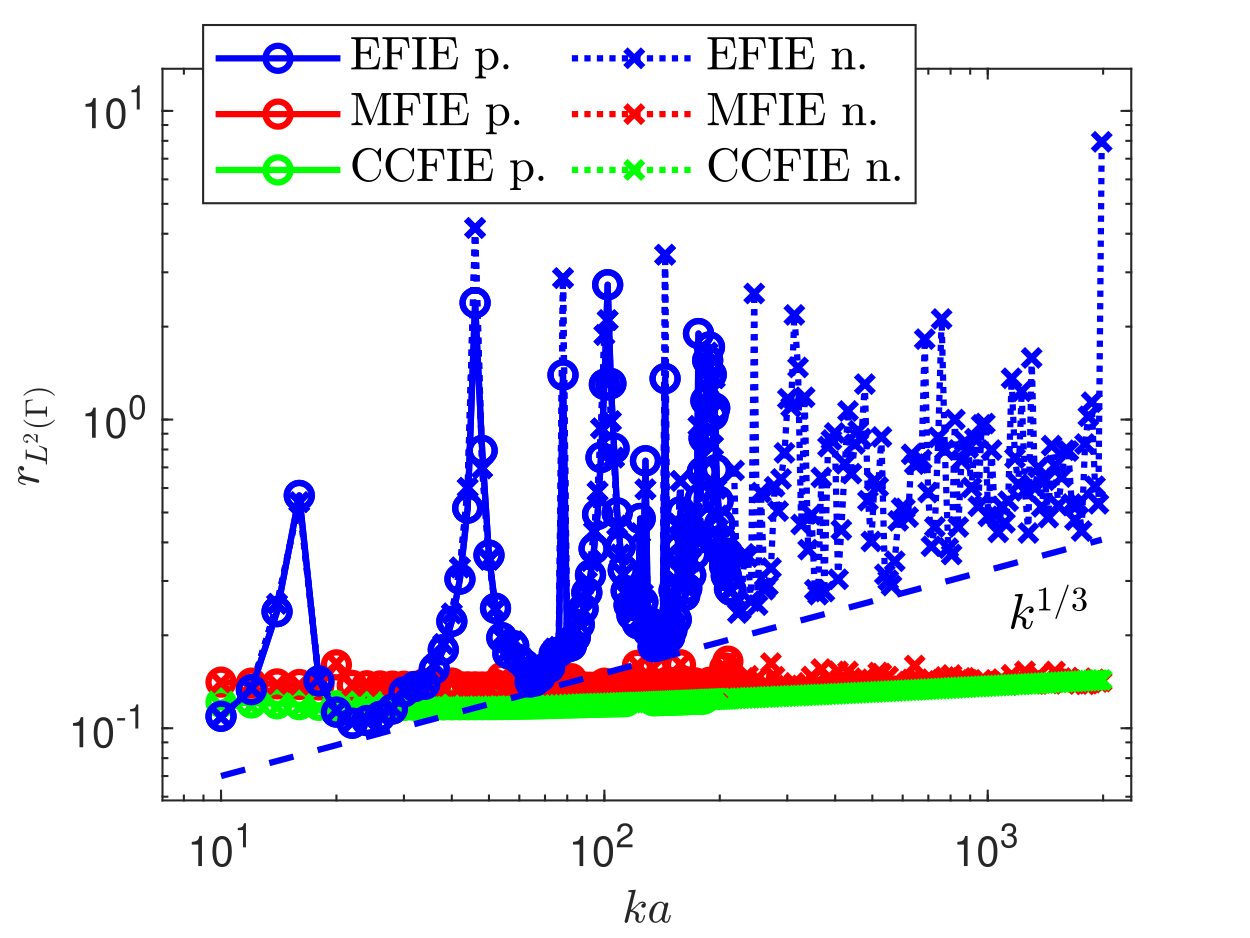}
}
\caption{Comparison between predicted (p.) and numerical (n.) results for the $L_2$ current error of the TM (a) and TE (b) formulations for varying $ka$ with $n_\lambda=4$.}
\label{fig:rl2}
\end{figure}

\begin{figure}
\centering
\subfloat[\label{subfig-1:}]{%
  \includegraphics[width=0.49
\columnwidth]{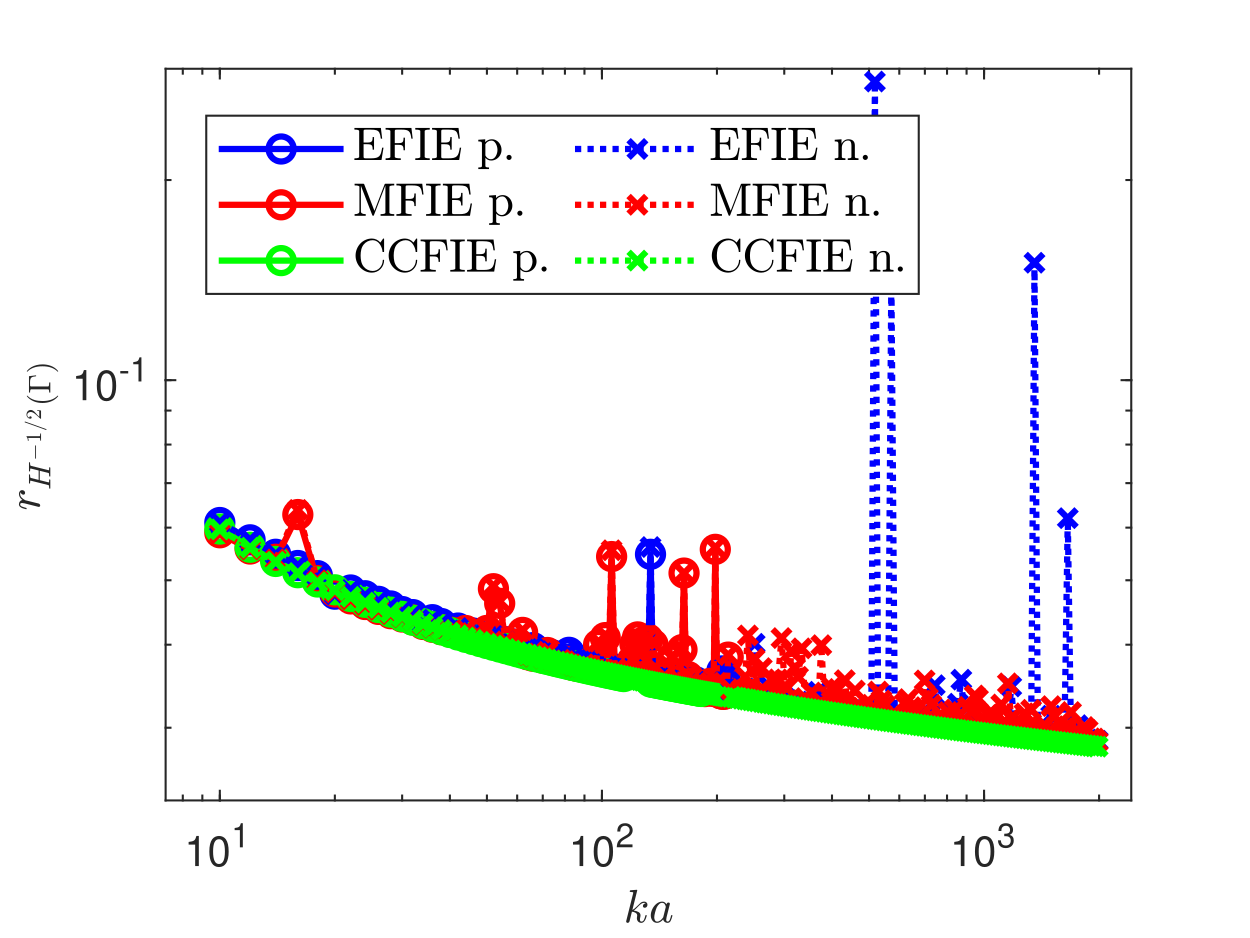}
}
\hfill
\subfloat[\label{subfig-2:}]{%
  \includegraphics[width=0.49\columnwidth]{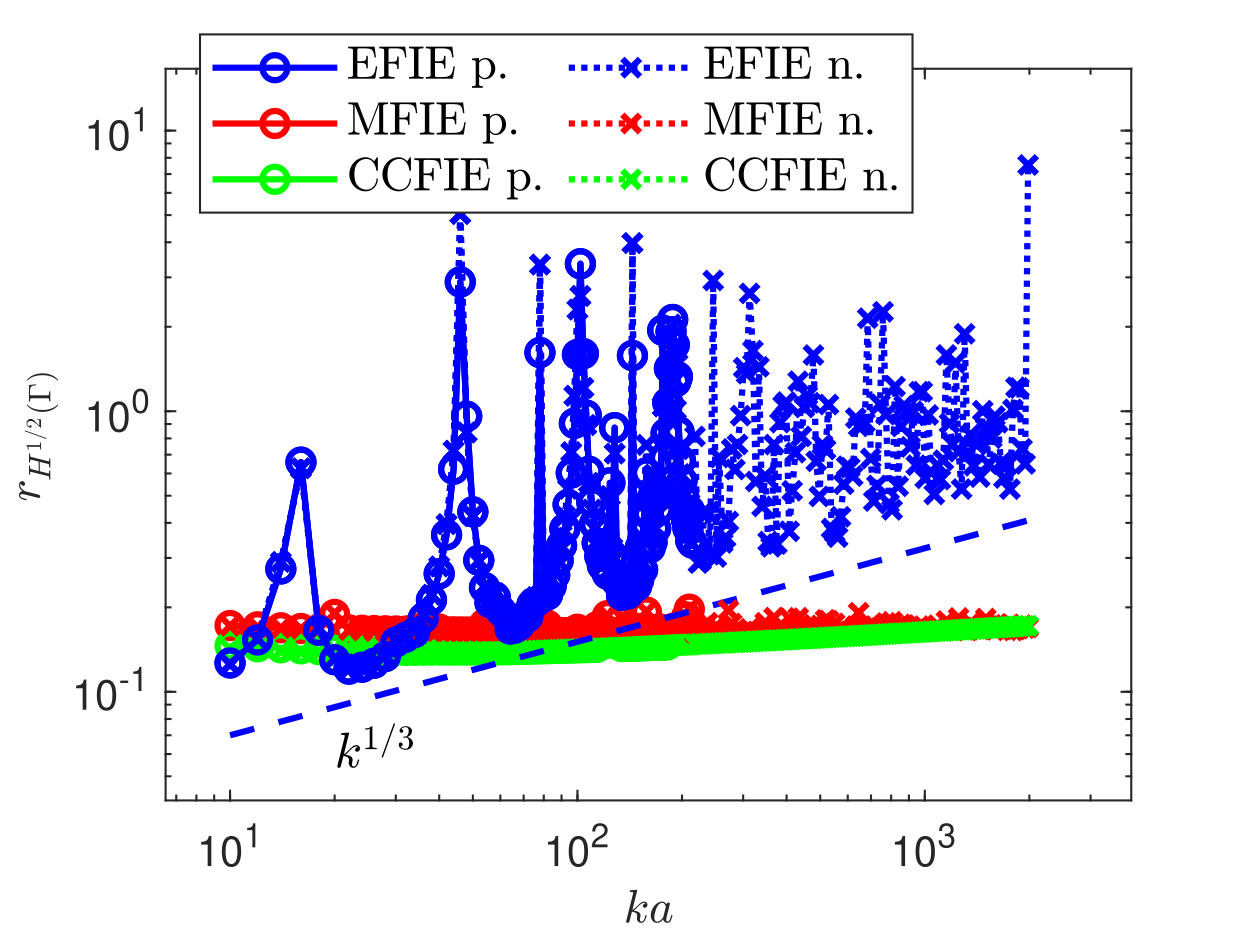}
}
\caption{Comparison between predicted (p.) and numerical (n.) results for the $H^s$ current error of the TM (a) and TE (b) formulations for varying $ka$ with $n_\lambda=4$.}
\label{fig:rhs}
\end{figure}

\begin{figure}
\centering
\subfloat[\label{subfig-1:}]{%
  \includegraphics[width=0.49
\columnwidth]{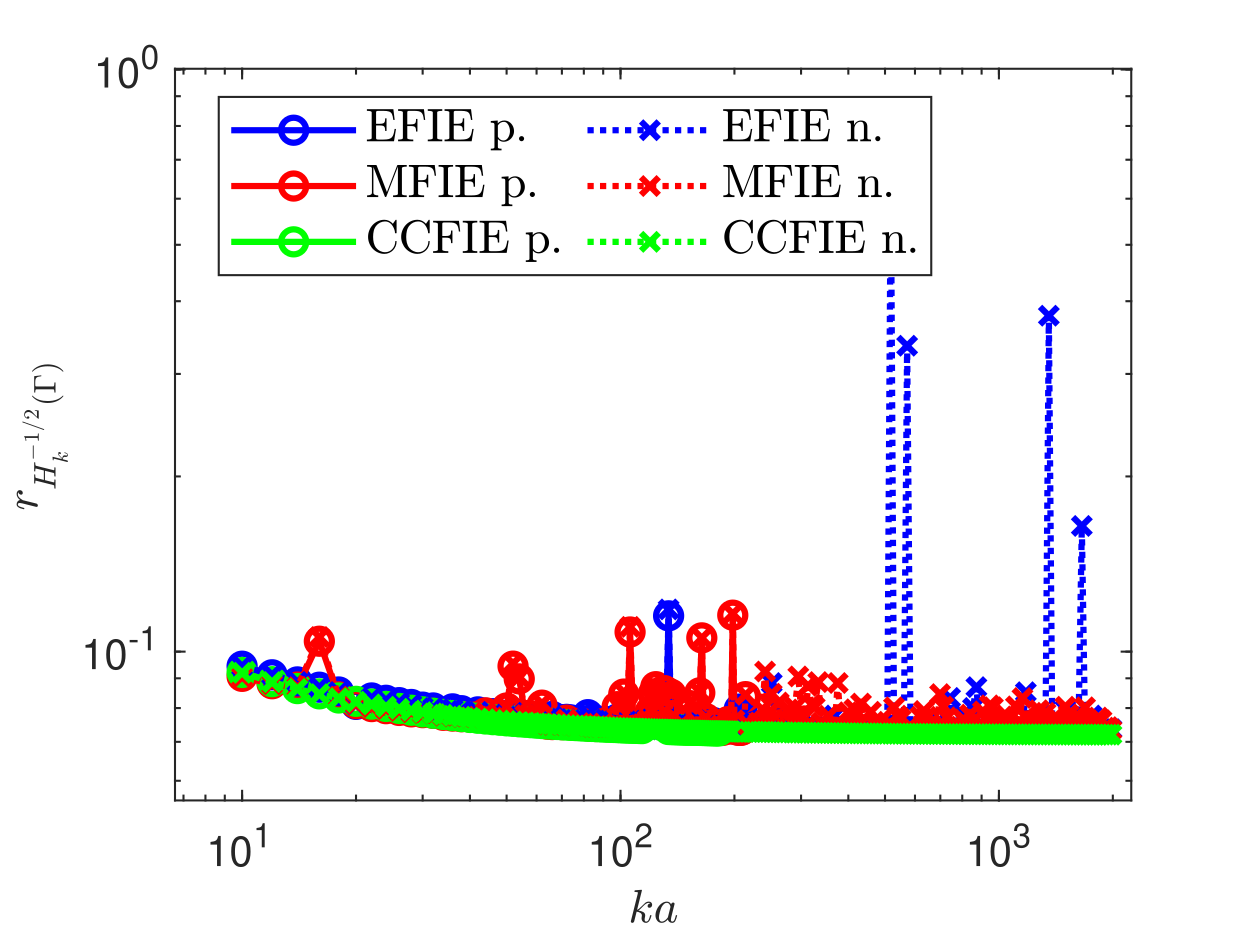}
}
\hfill
\subfloat[\label{subfig-2:}]{%
  \includegraphics[width=0.49\columnwidth]{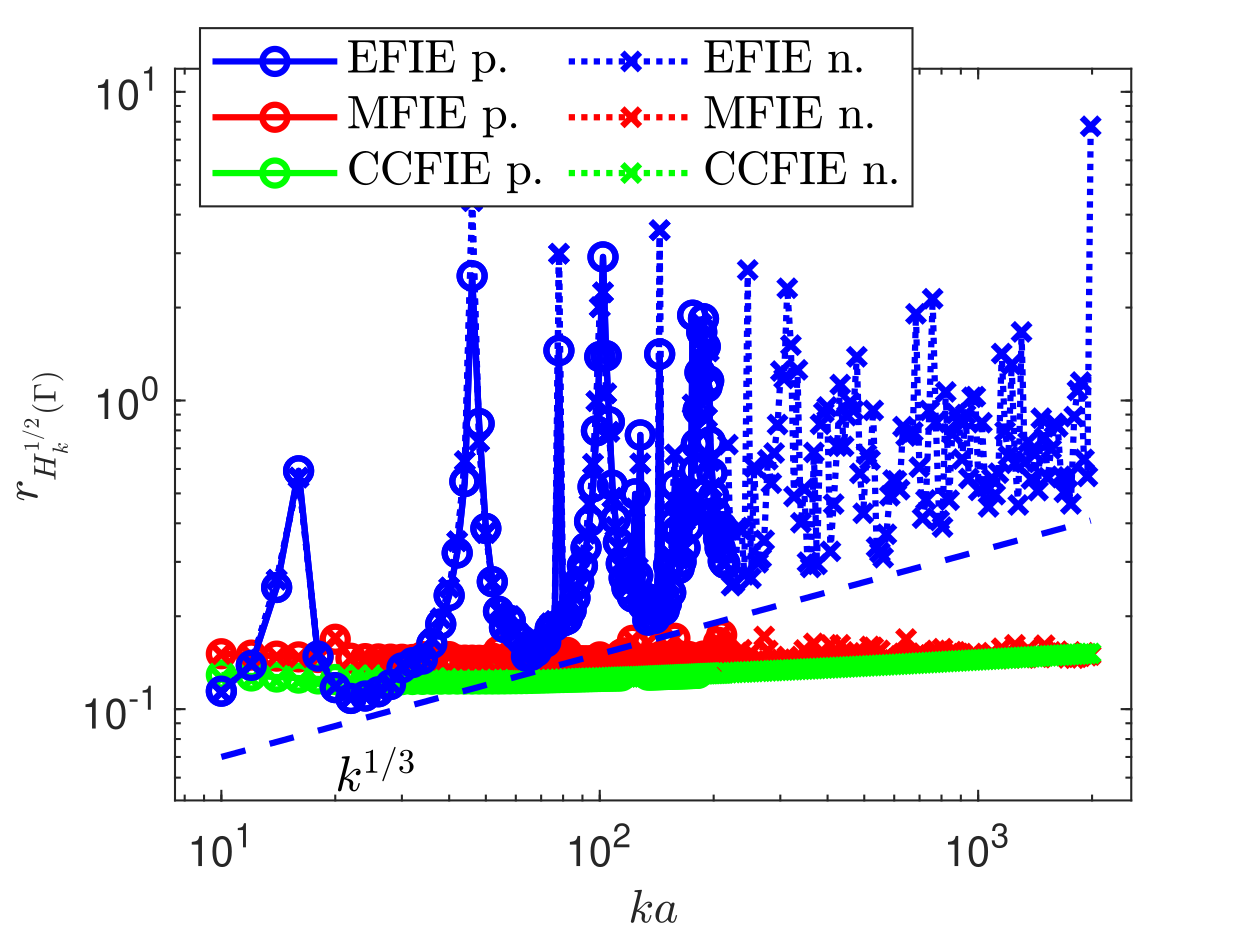}
}
\caption{Comparison between predicted (p.) and numerical (n.) results for the $H^s_k$ current error of the TM (a) and TE (b) formulations for varying $ka$ with $n_\lambda=4$.}
\label{fig:rhsk}
\end{figure}

\begin{figure}
\subfloat[\label{subfig-1:}]{%
  \includegraphics[width=0.45
\columnwidth]{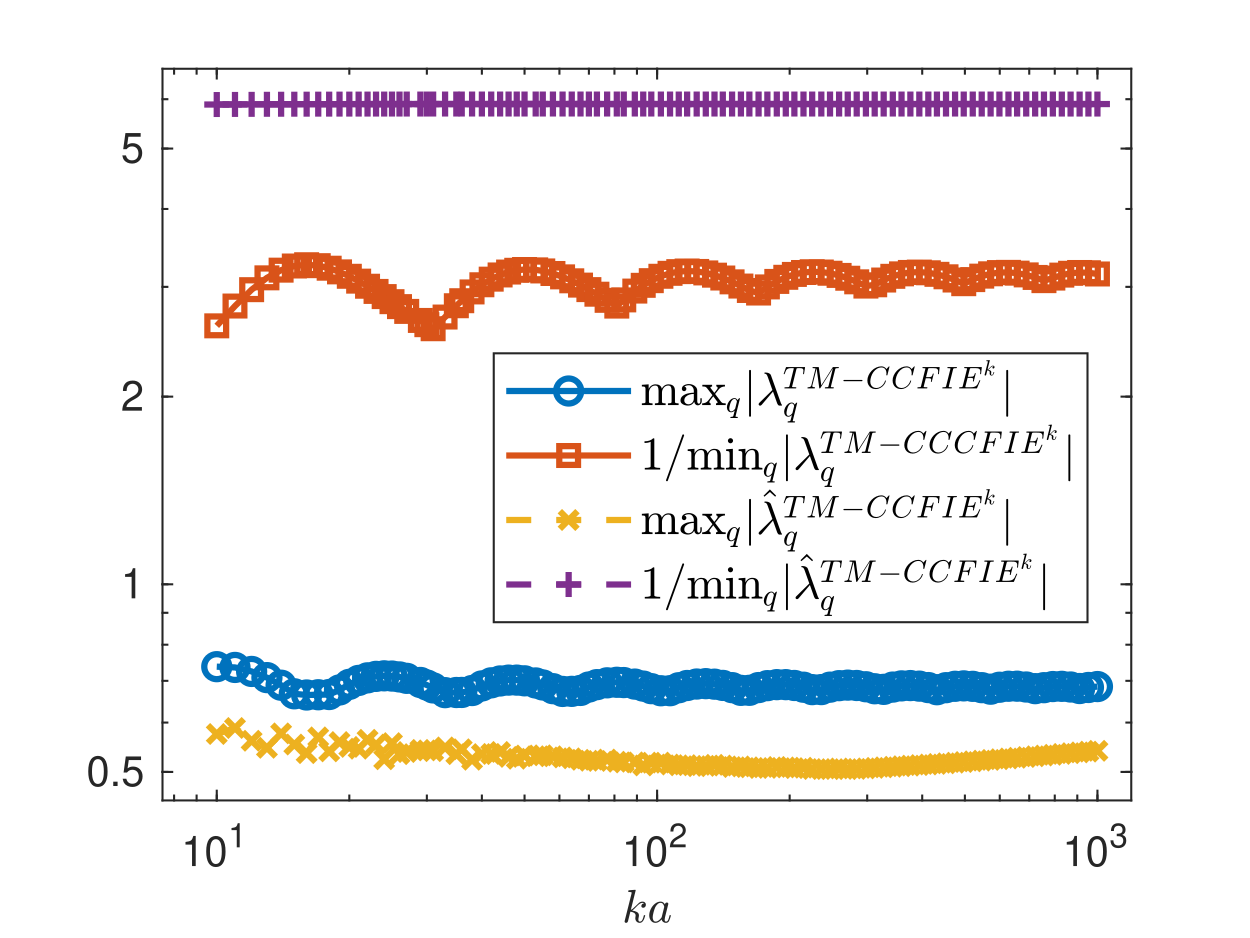}
}
\hfill
\subfloat[\label{subfig-2:}]{%
  \includegraphics[width=0.45\columnwidth]{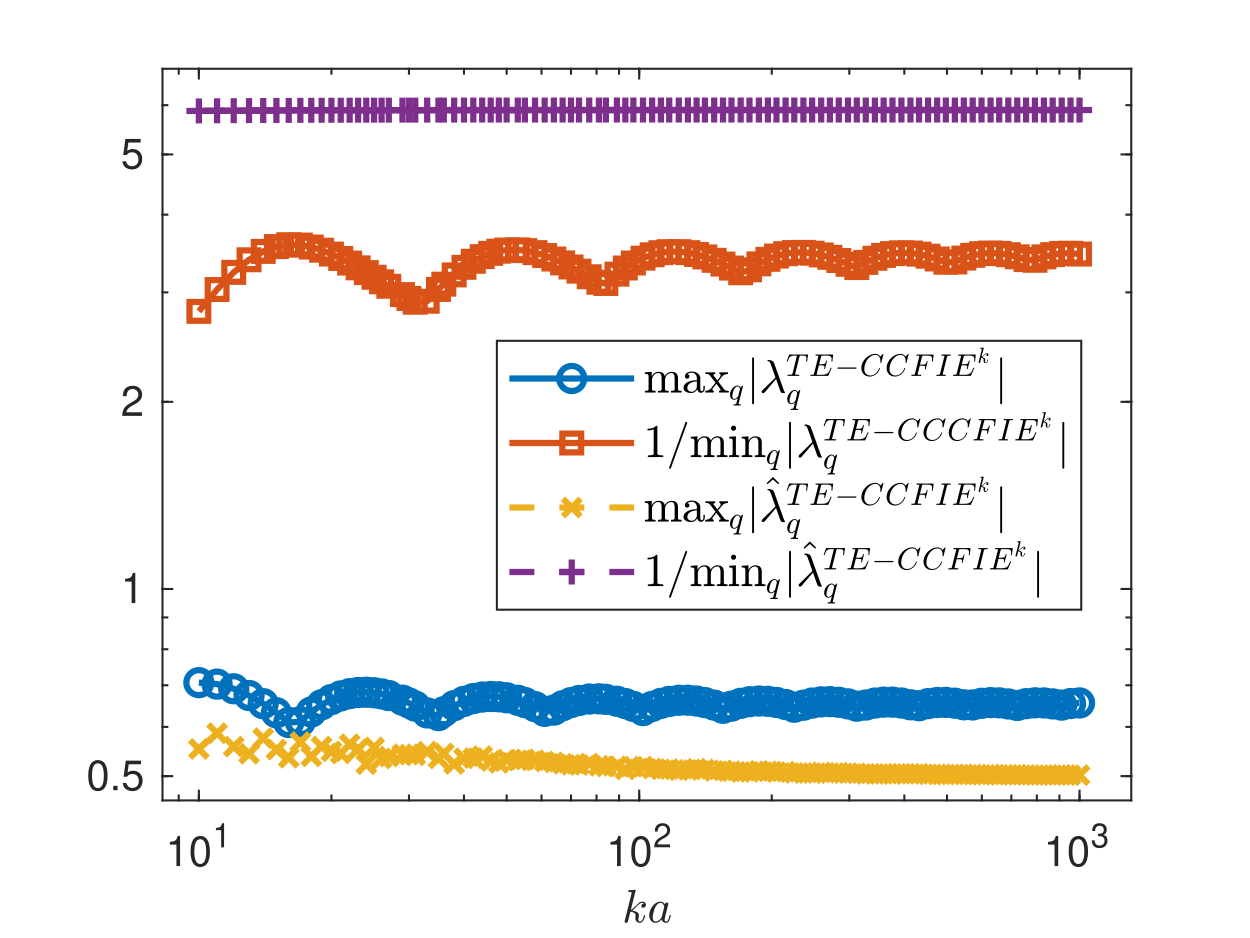}
}
\caption{$L_2$ norm of the CCFIO and its inverse, and $L_2$ norm of the CCFIO matrix evaluated with $n_\lambda = 4$ and its inverse, for the TM (a) and TE (b) formulations.}
\label{fig:CCFIO_wellcond}
\end{figure}

To validate the estimates on the current error presented in Section~\ref{sec:currenterror}, we first verify the convergence between our numerical results and predicted values for the diverse norms \eqref{eqn:rl2}, \eqref{eqn:rH}, \eqref{eqn:rHk} in \Cref{fig:rl2,fig:rhs,fig:rhsk}. All these scenarii show a good agreement between numerical and predicted results. The current error resulting from the solution of the TE-EFIE increases in the high-frequency regime as $(ka)^{1/3}$ away from resonances for all error measures, which is consistent with the theoretical expectations (Section~\ref{sec:erroranalysisHF}). The solution of the TE-CCFIE provides a current error that increases in frequency at a rate lower than $(ka)^{1/3}$. In particular this occurs in spite of the good conditioning properties of the  Calder\'{o}n combined field operators, summarized in \Cref{fig:CCFIO_wellcond}.

\subsection{Scattering error results}

Our models for the scattering error are compared to numerical results in \Cref{fig:sl2}. 
In accordance with theoretical expectations, the TM-EFIE provides a bounded scattering error, while both the TE-EFIE and the TE-CCFIE provide a scattering error that increases in the high-frequency regime. The increase rate is approximately equal to $(ka)^{1/9}$, compatible with the expectation of $\mathcal{O}((ka)^{1/3})$ (\Cref{sec:scaterroranalysisHF}). 

Moreover, as expected from \eqref{eqn:sTM} and \eqref{eqn:sTE}, in the MFIE case we observe sharp peaks of the scattering error at the resonance frequencies. We also notice an increase of $s_{L^2(\Gamma)}$ toward high frequencies away from resonances, not compatible with the theoretical expectation of bounded scattering error reported in \Cref{sec:scatteringerror}. We believe this may be due to the detrimental effect of spurious resonances.
On the other hand, the TE-EFIE also shows sharp error peaks at the resonant frequencies, while the TM-EFIE is more robust. 

\begin{figure}
\centering
\subfloat[\label{subfig-1:}]{%
  \includegraphics[width=0.49
\columnwidth]{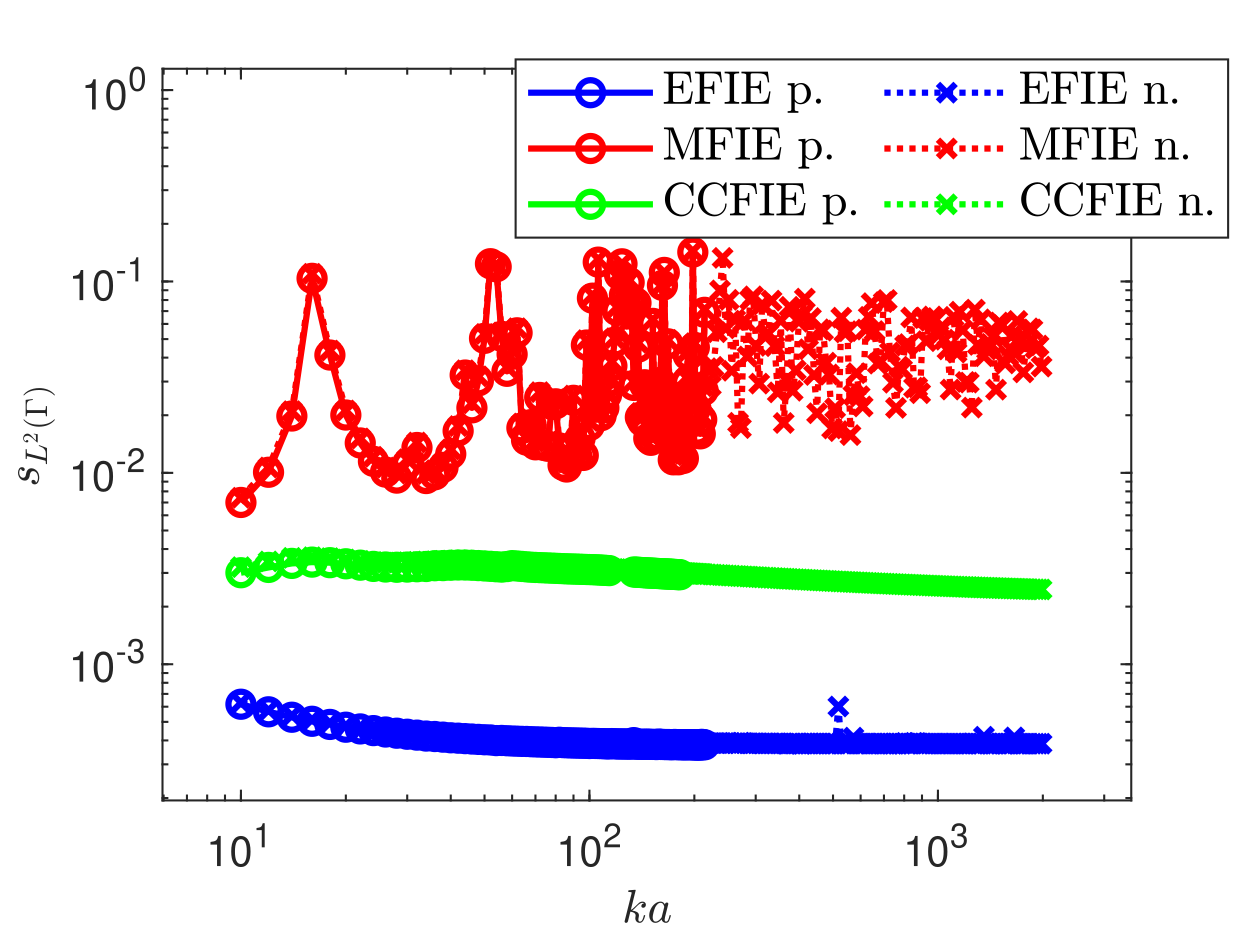}
}
\hfill
\subfloat[\label{subfig-2:}]{%
  \includegraphics[width=0.49\columnwidth]{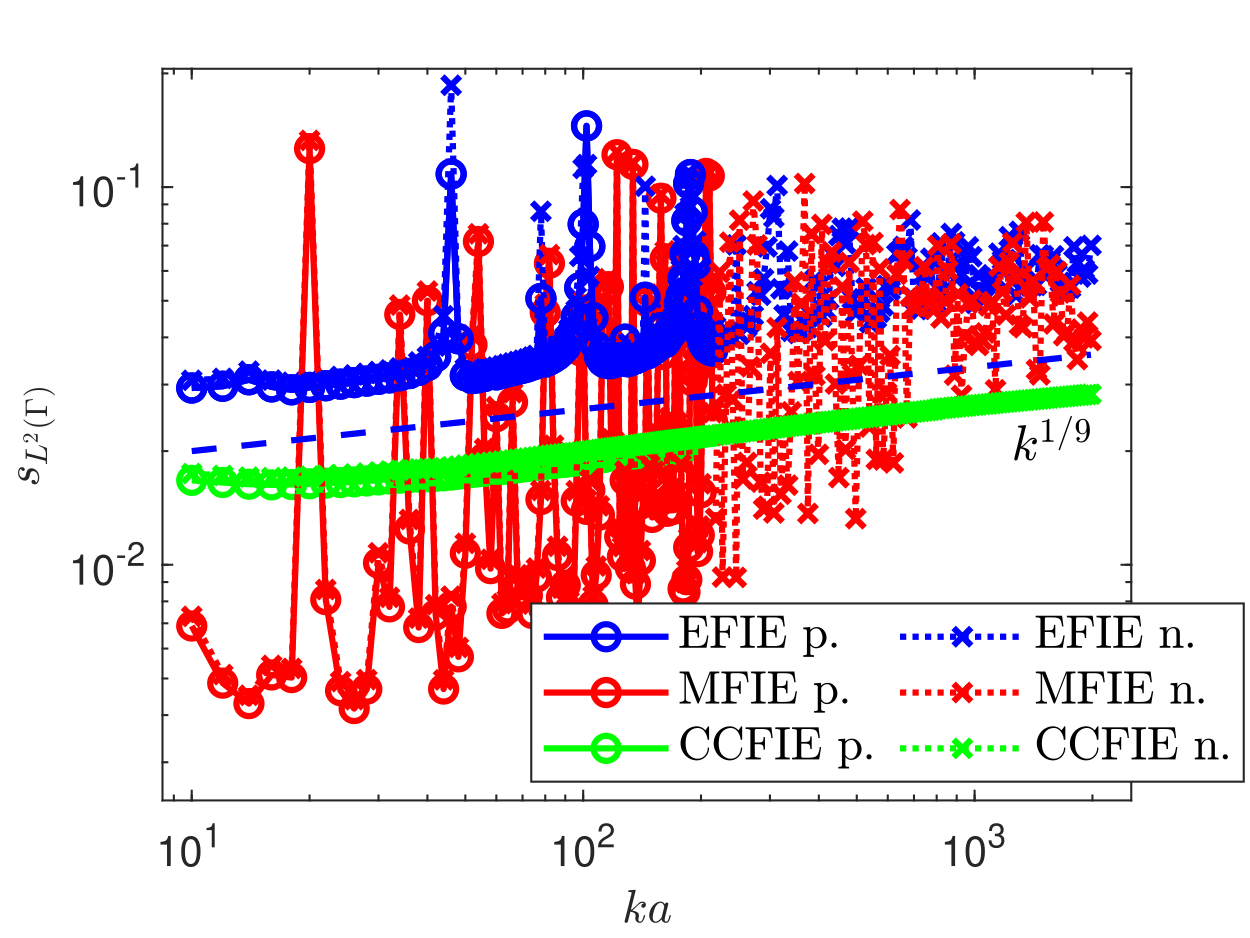}
}
\caption{Comparison between predicted (p.) and numerical (n.) results for the $L_2$ scattering error of the (a) and TE (b) formulations for varying $ka$ with $n_\lambda=4$.}
\label{fig:sl2}
\end{figure}

\subsection{Influence of the discretization parameter $n_\lambda$}
\begin{figure}
\centering
\subfloat[\label{subfig-1:rhsk_loh8_a}]{%
  \includegraphics[width=0.49
\columnwidth]{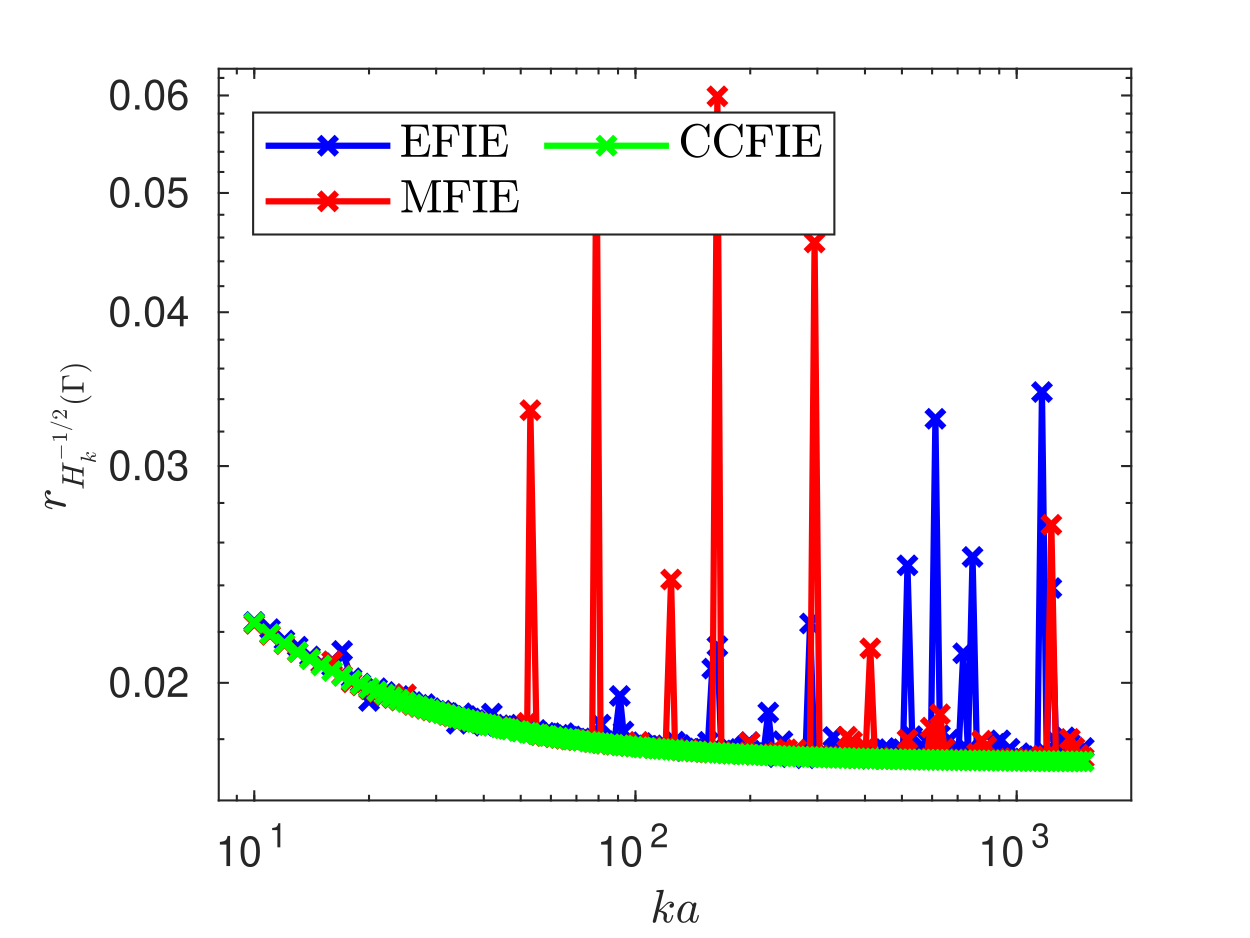}
}
\hfill
\subfloat[\label{subfig-2:rhsk_loh8_b}]{%
  \includegraphics[width=0.49\columnwidth]{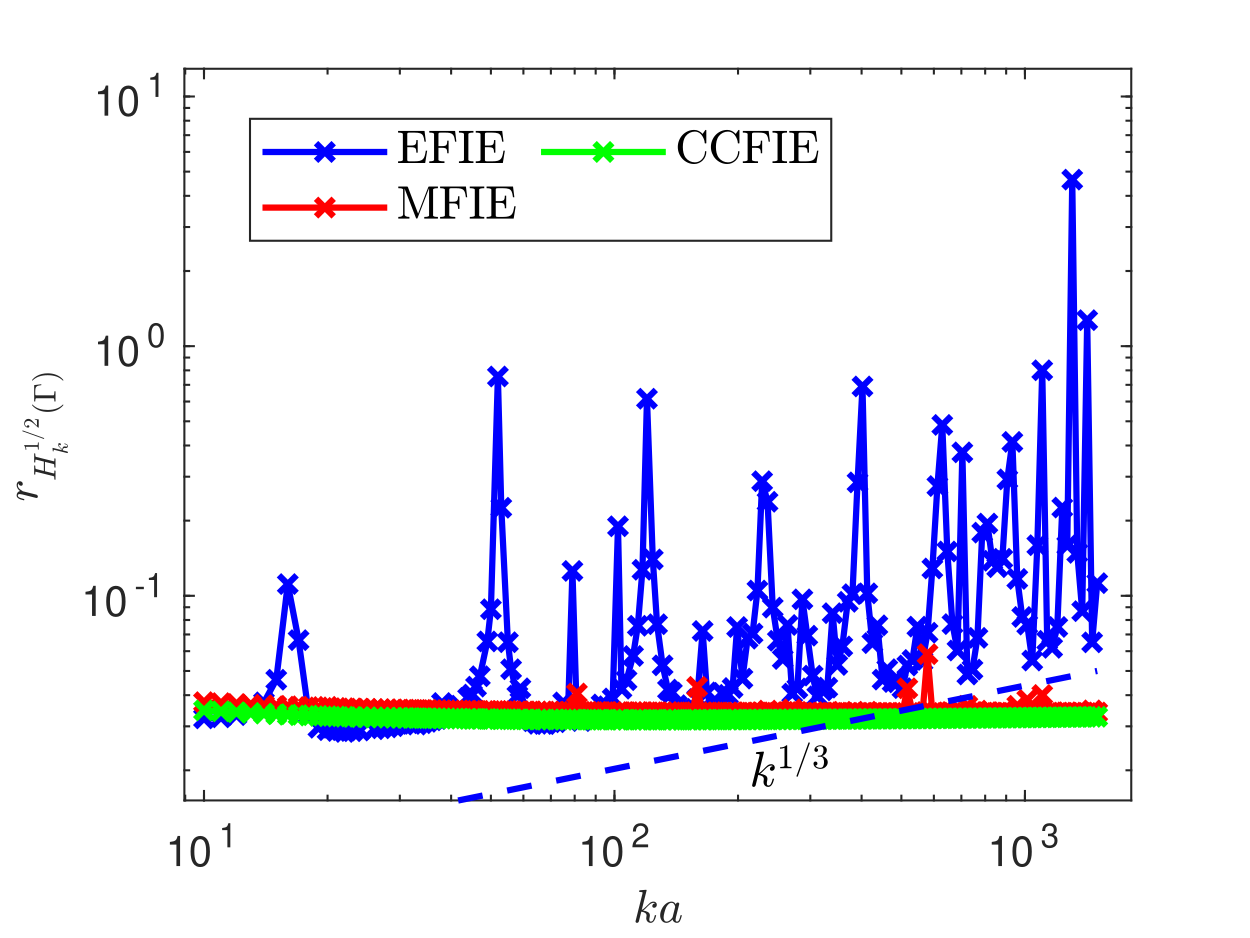}
}
\caption{Comparison between predicted (p.) and numerical (n.) results for the $H^s_k$ current error of the TM (a) and TE (b) formulations for varying $ka$ with $n_\lambda=8$.}
\label{fig:rhsk_loh8}
\end{figure}

\begin{figure}
\centering
\subfloat[\label{subfig-1:}]{%
  \includegraphics[width=0.49
\columnwidth]{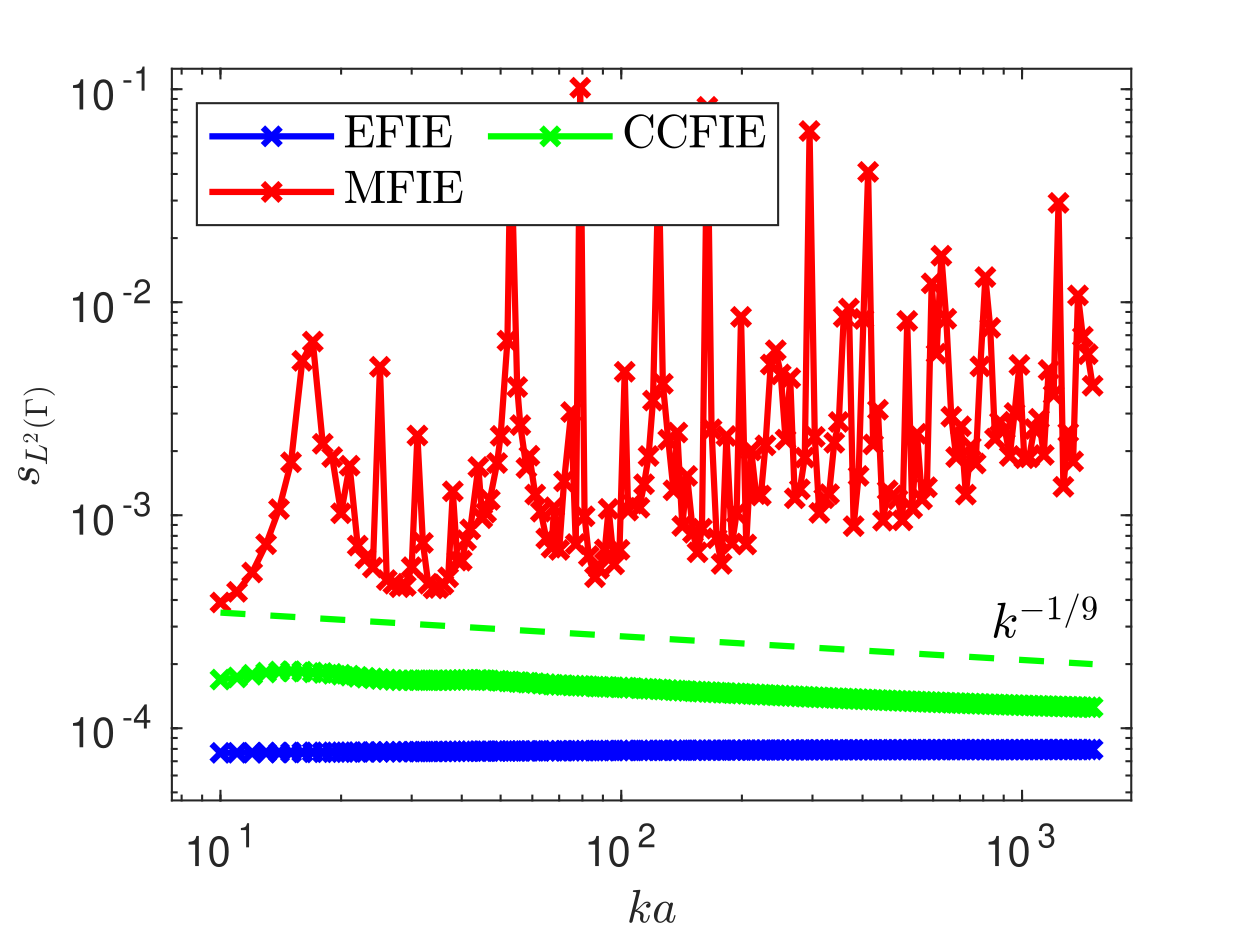}
}
\hfill
\subfloat[\label{subfig-2:}]{%
  \includegraphics[width=0.49\columnwidth]{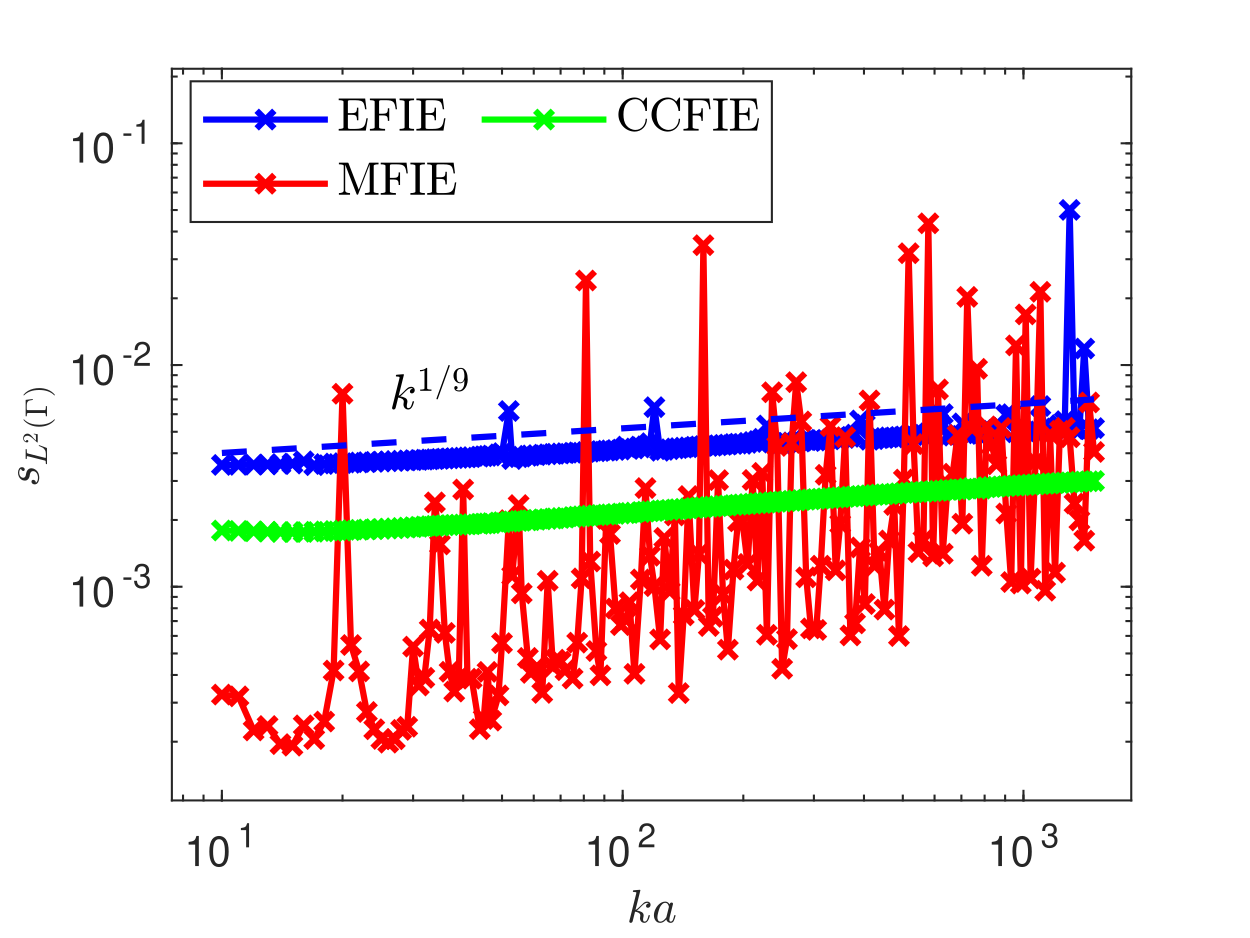}
}
\caption{Comparison between predicted (p.) and numerical (n.) results for the $L_2$ current error of the TM (a) and TE (b) formulations for varying $ka$ with $n_\lambda=8$.}
\label{fig:sl2_loh8}
\end{figure}

To investigate the influence of the discretization parameter $n_\lambda$ on the current and scattering errors, we numerically estimate these quantities under the condition $n_\lambda = 8$ (\Cref{fig:rhsk_loh8} and \Cref{fig:sl2_loh8}). Both the current and scattering error resulting from the BEM discretization at $n_\lambda = 8$ are significantly lower than the ones achieved at $n_\lambda = 4$. In \Cref{subfig-1:rhsk_loh8_a} we can still observe an asymptotic increase of the current error resulting from the TE-EFIE as $(ka)^{1/3}$ in the high-frequency regime, but in this case the onset of the increasing behavior, preceded by a constant behavior, is at an higher value of $ka$ than in the case of $n_\lambda = 4$. On the other hand, also the asymptotic increase of the scattering error from the TE-EFIE discretized at $n_\lambda = 8$ is comparable with the one at $n_\lambda = 4$, but in this case we do not observe a constant behavior of the scattering error at low frequencies.

\subsection{The proposed filtering approach}
\label{sec:numresults_filt}
\begin{figure}
\centering
\subfloat[\label{subfig-1:rl2_filt}]{%
  \includegraphics[width=0.49\columnwidth]{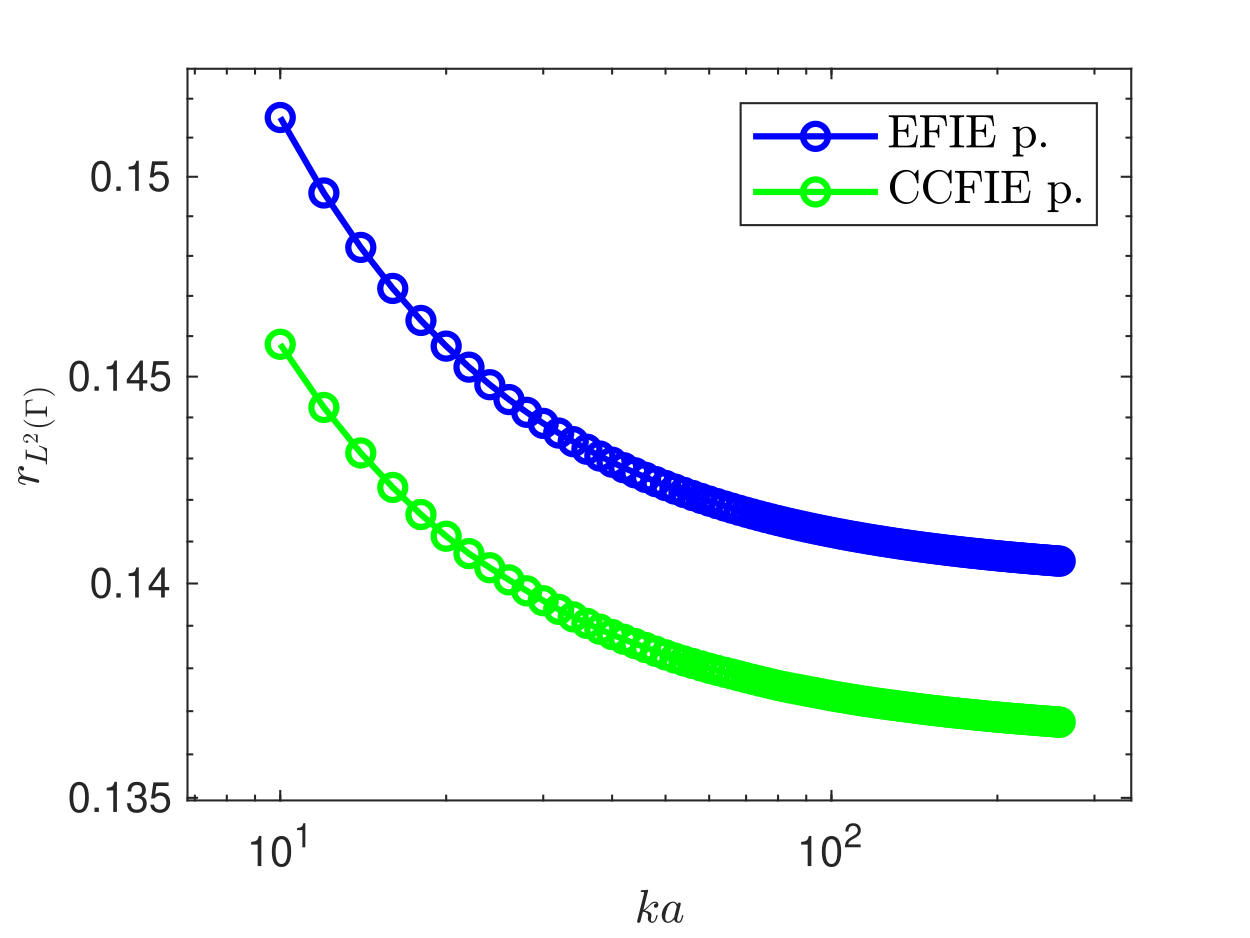}
}
\hfill
\subfloat[\label{subfig-2:rhs_filt}]{%
  \includegraphics[width=0.49\columnwidth]{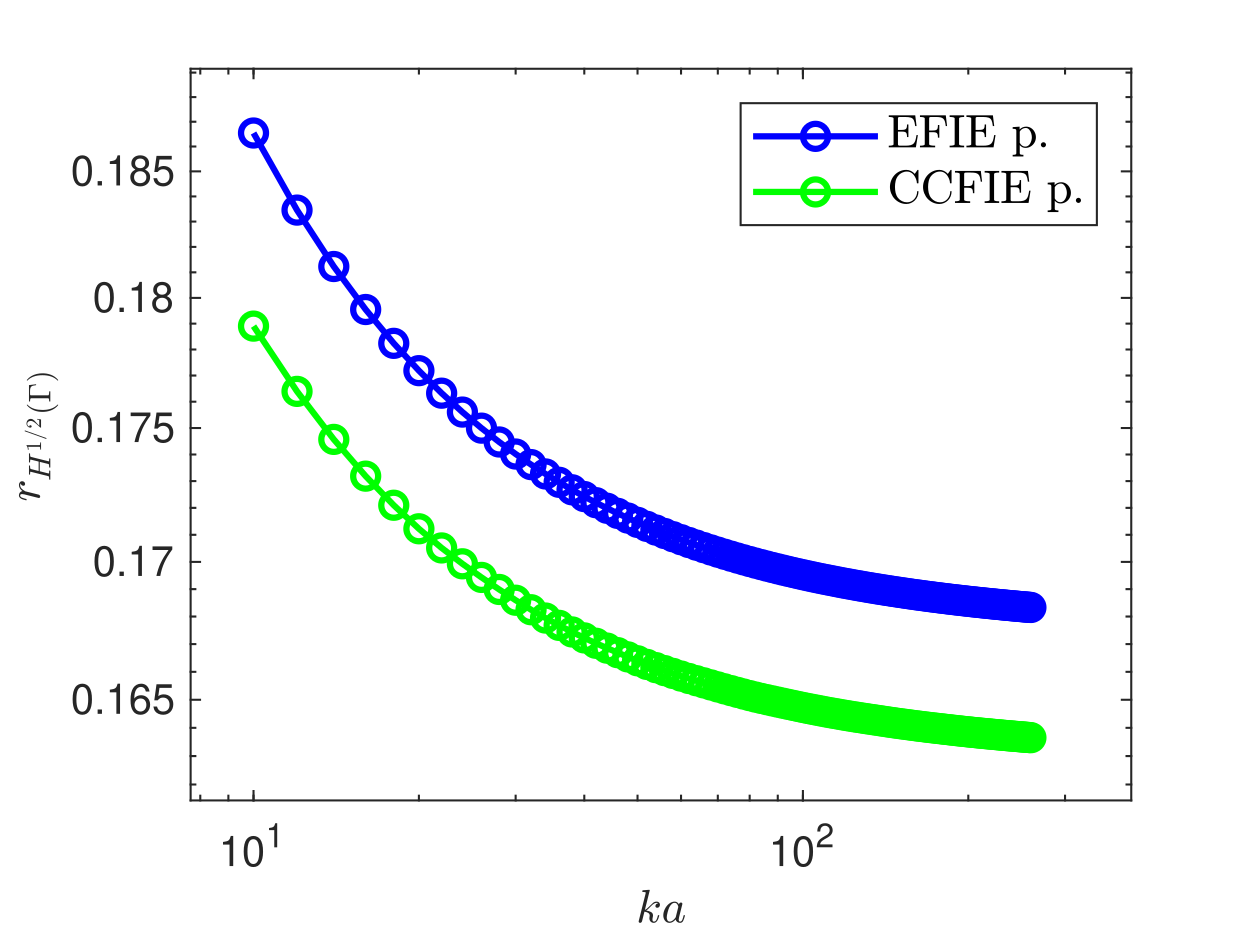}
}
\hfill
\subfloat[\label{subfig-3:rhsk_filt}]{%
  \includegraphics[width=0.49\columnwidth]{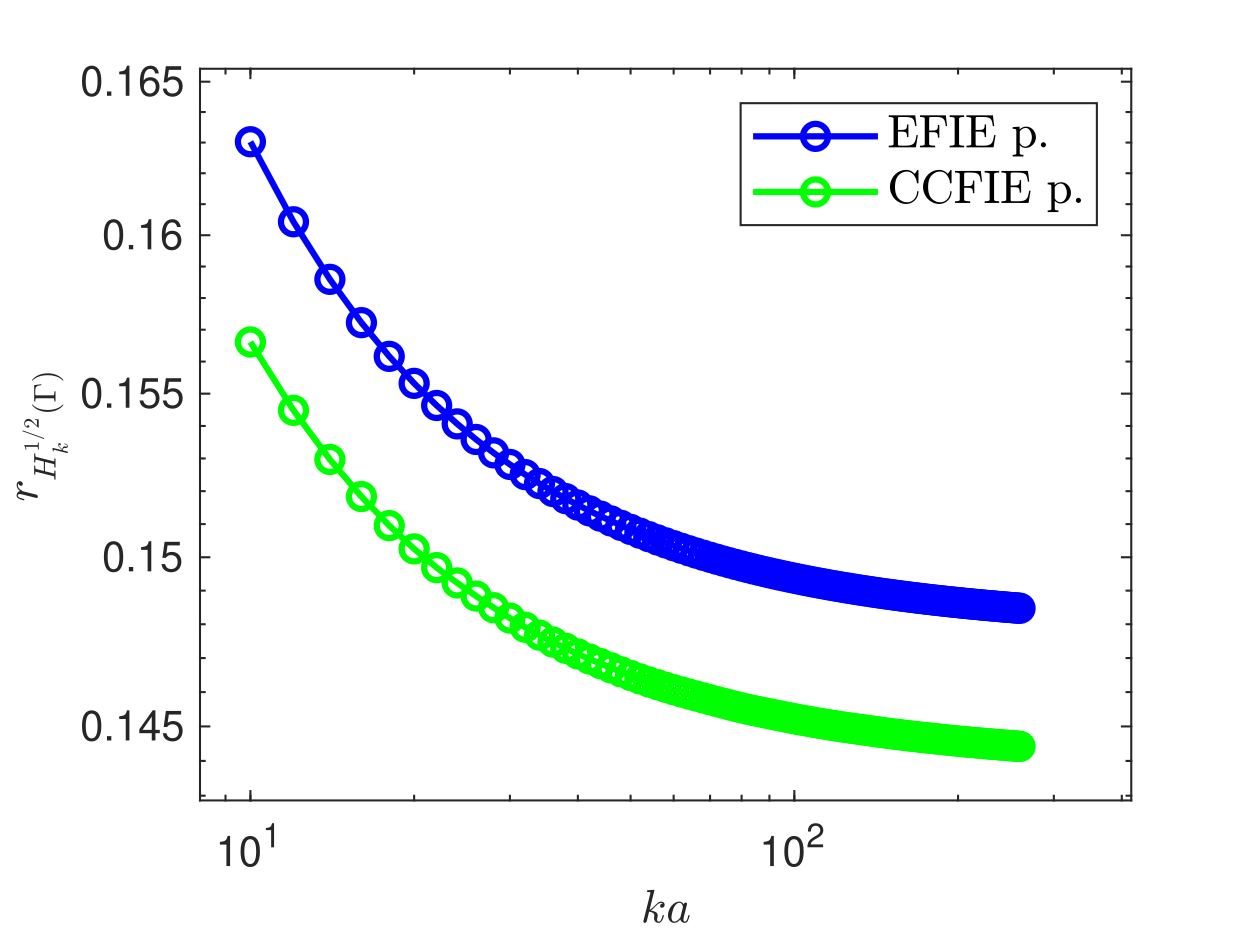}
}
\hfill
\subfloat[\label{subfig-4:sl2_filt}]{%
  \includegraphics[width=0.49\columnwidth]{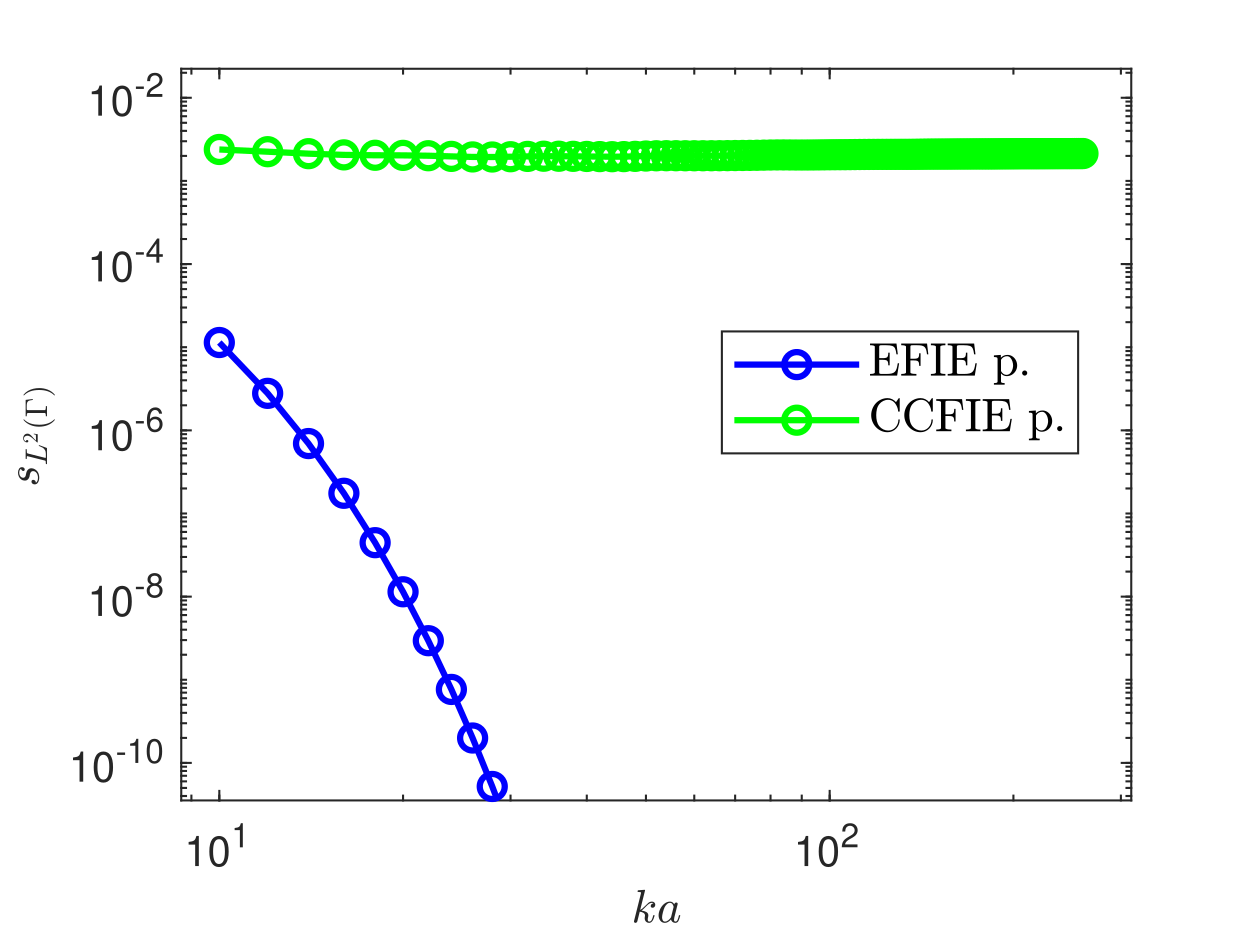}
}
\caption{$L_2$ (a), $H^s$ (b), $H^s_k$ (c) current error and $L_2$ scattering error for the TE-EFIE and TE-CCFIE built with the ideally filtered operator $\op N^k_F$ for varying $ka$ with $n_\lambda=4$.}
\label{fig:sl2}
\end{figure}

The proposed filtering approach has been applied to modify the TE-EFIE and TE-CCFIE by substituting the hypersingular operator $\op N^k$ with the ideally filtered operator $\op N_F^k$. The resulting current errors in the different measures are shown in \Cref{subfig-1:rl2_filt,subfig-2:rhs_filt,subfig-3:rhsk_filt} and the scattering error is represented in \Cref{subfig-4:sl2_filt}.
From these figures, we note that the current error of the filtered formulations does not increase in frequency. This was expected, as the spectral error of the discretization of $\op N_F^k$ and of all the operators involved in the $\text{TE-CCFIE}_F$ does not increase in the high-frequency regime.

Also, from the current error results, we immediately note the absence of the sharp increase of the current error in correspondence to resonance frequencies, as it is the case for the non-filtered formulations. This highlights the importance of the aliasing spectral error components over the resonant and quasi-resonant modes of the hypersingular operator in the hyperbolic region and its roles in determining the above mentioned peaks in the current error. The difference with respect to the TM-EFIE, resulting in milder peaks of current error in correspondence of resonance frequencies, is due to the fact that the eigenvalues of the hypersingular operators increase in modulo as $q$ toward high spectral indices, resulting in higher values of aliasing spectral error. 

As far as the scattering error---which is less prone to the projection spectral error than the current error---is concerned, since the coefficient $R_q$ decays in the elliptic region and since the error coefficient
\begin{equation}
    \rho_q^{\text{TE-EFIE}_F}=-\frac{E_q^{A,\op{N}_F^k}}{1+E_q^{\op{N}_F^k}}\,
\end{equation}
is exactly null in the hyperbolic and transition regions, the scattering error of the $\text{TE-EFIE}_F$ goes to zero, as shown in Figure \ref{subfig-4:sl2_filt}. On the other hand, the $\text{TE-CCFIE}_F$, differently from the TE-CCFIE, exhibits a bounded-in-frequency scattering error, stable at a value lower than the ones achieved by the non-filtered counterpart.

\section{Conclusions}
In this contribution, we analyzed several widely used boundary integral equations for the electromagnetic scattering problem from the canonical cylinder discretized via boundary element method. From the closed-form expressions for spectral, current, and scattering errors, we identified a form of pollution affecting some of them, that is, when increasing the frequency and keeping the product $kh$ constant (i.e., in the high-frequency regime), the spectral and solution errors increase. The proposed analysis has the power to clearly identify the roots of this phenomenon, and allows to take the first step toward its solution or mitigation. The study is corroborated by excellent agreement between theoretical findings and numerical results from our in-house BEM solver.

\appendix
\section{Derivation of the expression of $\upsilon_q^{\text{CCFIE}}$}
\label{sec:appendix}
Consider the TM-CCFIE (the derivation follows similarily for the TE-CCFIE),
\begin{equation}
    \left[{\op N^{\tilde{k}}} \op S^k  + \left(\frac{1}{2} \op I -  {\op D}^{\tilde{k}} \right)\left(\frac{1}{2} \op I +  \op D^{*k} \right)\right] (J_z) (\veg \rho)= \frac{1}{\mathrm{j}\eta}{\op N^{\tilde{k}}}E_z(\veg \rho)+\left(\frac{1}{2} \op I -  {\op D}^{\tilde{k}} \right)H_t(\veg \rho)\,.
\end{equation}
For ease of notation and without loss of generality, we assume the incident field is a plane-wave traveling along direction $\hat{\veg x}$.
Using cylindrical mode expansion of plane waves, the $n-$th elements of the arrays $\vec E_z$ and $\vec H_t$ \label{eqn:Fn} can be written as
\begin{align}
    \vec E_{z,n}&=\sum_{q=-\infty}^\infty \mathrm{j}^{-q} J_q(ka)F_{-q} e^{-\mathrm{j}q\phi_n}\,,\\
    \vec H_{t,n}&=\frac{1}{\mathrm{j}\eta}\sum_{q=-\infty}^\infty \mathrm{j}^{-q} J_q^\prime(ka)F_{-q} e^{-\mathrm{j}q\phi_n}\,.
\end{align}
Hence, the $q-$th Fourier coefficient of the array $\hat{\vec J}_z$ resulting from the discretization of the integral formulation is
\begin{equation}
    \hat{U}_q^{\text{TM}} = \frac{\frac{\hat{\lambda}^{\op N^{\tilde{k}}}(\hat{\lambda}^{\mathcal{I}})^{-1}}{\mathrm{j}\eta}\mathrm{j}^{-q} J_q(ka)F_{-q}+
    \frac{\hat{\lambda}^{\text{TE-MFIO}^{\tilde{k}}}(\hat{\lambda}^{\mathcal{I}})^{-1}}{\mathrm{j}\eta}\mathrm{j}^{-q} J_q^\prime(ka)F_{-q}
    }{\hat{\lambda}^{\op N^{\tilde{k}}}(\hat{\lambda}^{\mathcal{I}})^{-1}\hat{\lambda}^{\op S^{{k}}}+\hat{\lambda}^{\text{TE-MFIO}^{\tilde{k}}}(\hat{\lambda}^{\mathcal{I}})^{-1}\hat{\lambda}^{\text{TM-MFIO}^{{k}}}}\,.
\end{equation}
Conversely, the $q-$th Fourier coefficient of the solution of the continuous integral formulation is $U_q^{\text{TM}}$ \eqref{eqn:UTM}, which can be written as
\begin{equation}
    U_q^{\text{TM}} = \frac{\frac{{\lambda}^{\op N^{\tilde{k}}}}{\mathrm{j}\eta}\mathrm{j}^{-q} J_q(ka)+
    \frac{{\lambda}^{\text{TE-MFIO}^{\tilde{k}}}}{\mathrm{j}\eta}\mathrm{j}^{-q} J_q^\prime(ka)
    }{{\lambda}^{\op N^{\tilde{k}}}{\lambda}^{\op S^{{k}}}+{\lambda}^{\text{TE-MFIO}^{\tilde{k}}}{\lambda}^{\text{TM-MFIO}^{{k}}}}\,.
\end{equation}
It follows that the coefficient $\upsilon_q^{\text{TM-CCFIE}}=(\hat{U}_q^{\text{TM}}-U_q^{\text{TM}})/U_q^{\text{TM}}$ reads
\begin{align}
    \upsilon_q^{\text{TM-CCFIE}} = \, &F_{-q}
    \frac{J_q(ka)\hat{\lambda}^{\op N^{\tilde{k}}}(\hat{\lambda}^{\mathcal{I}})^{-1}+J_q^\prime(ka)\hat{\lambda}^{\text{TE-MFIO}^{\tilde{k}}}(\hat{\lambda}^{\mathcal{I}})^{-1}}{J_q(ka){\lambda}^{\op N^{\tilde{k}}}+J_q^\prime(ka){\lambda}^{\text{TE-MFIO}^{\tilde{k}}}}\cdot\nonumber\\
    &\frac{\lambda_q^{\text{TM-CEFIO}}+\lambda_q^{\text{TM-CMFIO}}}{\lambda_q^{\text{TM-CEFIO}}E_q^{\text{TM-CEFIO}}+\lambda_q^{\text{TM-CMFIO}}E_q^{\text{TM-CMFIO}}}
    -1\,.
\end{align}
After noticing that
\begin{equation}
    \frac{J_q(ka)\hat{\lambda}^{\op N^{\tilde{k}}}(\hat{\lambda}^{\mathcal{I}})^{-1}+J_q^\prime(ka)\hat{\lambda}^{\text{TE-MFIO}^{\tilde{k}}}(\hat{\lambda}^{\mathcal{I}})^{-1}}{J_q(ka){\lambda}^{\op N^{\tilde{k}}}+J_q^\prime(ka){\lambda}^{\text{TE-MFIO}^{\tilde{k}}}}=
    \frac{\frac{1+E_q^{\op N^{\tilde{k}}}}{1+E_q^{\op I}} J_q^\prime(\tilde{k}a)J_q(ka)+\frac{1+E_q^{\text{TE-MFIO}^{\tilde{k}}}}{1+E_q^{\op I}} J_q(\tilde{k}a)J_q^\prime(ka)}
    {J_q^\prime(\tilde{k}a)J_q(ka)+J_q(\tilde{k}a)J_q^\prime(ka)}\,,
\end{equation}
as a result of cancellation of the contribution $H_q^{(2)\prime}(\tilde{k}a)$, and that
\begin{align}
    &\frac{\lambda_q^{\text{TM-CEFIO}}+\lambda_q^{\text{TM-CMFIO}}}{\lambda_q^{\text{TM-CEFIO}}E_q^{\text{TM-CEFIO}}+\lambda_q^{\text{TM-CMFIO}}E_q^{\text{TM-CMFIO}}}   =\left(J_q^\prime(\tilde{k}a)J_q(ka)+J_q(\tilde{k}a)J_q^\prime(ka)\right) \nonumber\\
    &\bigg(\frac{(1+E_q^{\op N^{\tilde{k}}})(1+E_q^{\op S^{{k}}})}{(1+E_q^{\op I})}J_q^\prime(\tilde{k}a)J_q(ka) + \frac{(1+E_q^{\text{TE-MFIO}^{\tilde{k}}})(1+E_q^{\text{TM-MFIO}^{{k}}})}{(1+E_q^{\op I})}J_q(\tilde{k}a)J_q^\prime(ka)\bigg)^{-1}\,,
\end{align}
which results from the cancellation of the product $H_q^{(2)\prime}(\tilde{k}a)H_q^{(2)}({k}a)$, $\upsilon_q^{\text{TM-CCFIE}}$ can be written as, 
\begin{align} 
\upsilon_q^{\text{TM-CCFIE}}&=F_{-q}
    \frac{({1+E_q^{\op N^{\tilde{k}}}}) J_q^\prime(\tilde{k}a)J_q(ka)+({1+E_q^{\text{TE-MFIO}^{\tilde{k}}}}) J_q(\tilde{k}a)J_q^\prime(ka)}
    {{(1+E_q^{\op N^{\tilde{k}}})(1+E_q^{\op S^{{k}}})}J_q^\prime(\tilde{k}a)J_q(ka) + {(1+E_q^{\text{TE-MFIO}^{\tilde{k}}})(1+E_q^{\text{TM-MFIO}^{{k}}})}J_q(\tilde{k}a)J_q^\prime(ka)}-1\\
    &=\frac{J_q^\prime(\tilde{k}a)J_q(ka)(1+E_q^{\op N^{\tilde{k}}})(1+E_q^{\op S^{{k}}})}{J_q^\prime(\tilde{k}a)J_q(ka)(1+E_q^{\op N^{\tilde{k}}})(1+E_q^{\op S^{{k}}})+J_q(\tilde{k}a)J_q^\prime(ka)(1+E_q^{\text{TE-MFIO}^{\tilde{k}}})(1+E_q^{\text{TM-MFIO}^{{k}}})}\left(\frac{F_{-q}}{(1+E_q^{\op S^{{k}}})}-1\right)\nonumber\\
    &+\frac{J_q(\tilde{k}a)J_q^\prime(ka)(1+E_q^{\text{TE-MFIO}^{\tilde{k}}})(1+E_q^{\text{TM-MFIO}^{{k}}})}{J_q^\prime(\tilde{k}a)J_q(ka)(1+E_q^{\op N^{\tilde{k}}})(1+E_q^{\op S^{{k}}})+J_q(\tilde{k}a)J_q^\prime(ka)(1+E_q^{\text{TE-MFIO}^{\tilde{k}}})(1+E_q^{\text{TM-MFIO}^{{k}}})}\left(\frac{F_{-q}}{(1+E_q^{\text{TM-MFIO}^{{k}}})}-1\right)
    \label{eqn:upsAppendix}
\end{align}
which is equal to
\begin{equation}
    \upsilon_q^{\text{TM-CCFIE}}=\frac{\hat{\lambda}_q^{\text{TM-CEFIO}}\upsilon_q^{\text{TM-EFIE}}+\hat{\lambda}_q^{\text{TM-CMFIO}}\upsilon_q^{\text{TM-MFIE}}}{\hat{\lambda}_q^{\text{TM-CEFIO}}+\hat{\lambda}_q^{\text{TM-CMFIO}}}\,.
\end{equation}

\section*{Acknowledgment}
This work was supported by the European Innovation Council (EIC) through the European Union’s Horizon Europe research Programme under Grant 101046748 (Project CEREBRO) and by the European Union – Next Generation EU within the PNRR project ``Multiscale modeling and Engineering Applications'' of the Italian National Center for HPC, Big Data and Quantum Computing (Spoke 6) – PNRR M4C2, Investimento 1.4 - Avviso n. 3138 del 16/12/2021 - CN00000013 National Centre for HPC, Big Data and Quantum Computing (HPC) - CUP E13C22000990001.

\bibliographystyle{plainnat}

\bibliography{private_vgiunzioni.bib}

\end{document}